# Analytical investigations of quasi-circular frozen orbits in the Martian gravity field

Xiaodong Liu[1], Hexi Baoyin[2], and Xingrui Ma[3]

*School of Aerospace, Tsinghua University, Beijing 100084 CHINA*

*Email*: *liu-xd08@mails.tsinghua.edu.cn; baoyin@tsinghua.edu.cn; maxr@spacechina.com*

## Abstract

Frozen orbits are always important foci of orbit design because of their valuable characteristics that their eccentricity and argument of pericentre remain constant on average. This study investigates quasi-circular frozen orbits and examines their basic nature analytically using two different methods. First, an analytical method based on Lagrangian formulations is applied to obtain constraint conditions for Martian frozen orbits. Second, Lie transforms are employed to locate these orbits accurately, and draw the contours of the Hamiltonian to show evolutions of the equilibria. Both methods are verified by numerical integrations in an $80\times80$ Mars gravity field. The simulations demonstrate that these two analytical methods can provide accurate enough results. By comparison, the two methods are found well consistent with each other, and both discover four families of Martian frozen orbits: three families with small eccentricities and one family near the critical inclination. The results also show some valuable conclusions: for the majority of Martian frozen orbits, argument of pericentre are kept at 270 degrees because $J_3$ has the same sign with $J_2$; while for a minority of ones with low altitude and low inclination, argument of pericentre are possible to be kept at 90 degrees because of the effect of the higher degree odd zonals; for the critical inclinations cases, argument of pericentre can also be kept at 90 degrees. It is worthwhile to note that there exist some special frozen orbits with extremely small eccentricity, which could provide much convenience for reconnaissance. Finally, the stability of Martian frozen orbits is estimated based on the trace of the monodromy matrix. The analytical investigations can provide good initial conditions for numerical correction methods in the more complex models.

[1] PhD candidate, School of Aerospace, Tsinghua University

[2] Associate Professor, School of Aerospace, Tsinghua University

[3] Professor, School of Aerospace, Tsinghua University

## 1 Introduction

Recently, along with the new wave of deep space exploration, Mars has aroused people's great curiosity once again because of its strong resemblance to Earth. For some missions conducted around Mars, the application of frozen orbits may be a good choice. The eccentricity and argument of pericentre of these orbits can remain constant on average, which is of great advantage to Martian missions. Extensive research has been conducted on frozen orbits about Earth and the Moon. However, so far there is a little previous literature on the study of Martian frozen orbits.

The notion of *frozen orbits* dates back to many years ago, the history of which is well explained in (Coffey et al. 1994). Many investigators have contributed research. An analytical method based on the canonical relationship of the partial derivatives of the Delaunay variables with respect to the gravitational disturbing potential was employed to predict the long-term evolution of the eccentricity and argument of pericentre without numerical integration (Smith 1986). Based on the Lagrange's planetary equations, the constraint equation for Earth frozen orbits can be derived for the gravity model involving higher degree zonals (Rosborough and Ocampo 1991). By Lie transformation to obtain the normalized Hamiltonian up to $J_9$, families of frozen orbits around the Earth-like planet can be obtained (Coffey et al. 1994), and the same method has been applied to find frozen orbits for lunar orbiter (Abad et al. 2009). With the polar component of the angular momentum as the parameter, natural families of Earth frozen orbits were continued numerically based on variational equations in the zonal gravity field up to $J_7$ (Lara et al. 1995), and this method was also applied to lunar frozen orbits (Elipe and Lara 2003). Using epicycle description to avoid the singularity of zero eccentricity and the critical inclination (the critical inclination is so-called because it is an intrinsic singularity in the artificial satellite theory (Coffey et al. 1986), and the value of it is 63.43°), analytical computation for Earth frozen orbits can be extended to arbitrary terms of zonal harmonics (Aorpimai and Palmer 2003).

For near-circular lunar mapping orbits, numerical investigations were used to find initial conditions to improve the orbital lifetime and reduce the maneuver for stationkeeping (Ramanan and Adimurthy 2005). Analytical formulation and numerical simulations combined with a traditional differential correction (DC) process were ever used to select usable lunar frozen orbits in the full gravity field (Folta and Quinn 2006). The long-lifetime lunar repeating ground track orbits were characterized in the Earth-Moon restricted three-body problem perturbed by a high-resolution lunar gravity field (Russell and Lara 2007). Families of circulating eccentric orbits around planetary moons were calculated in the averaged problem via contour plots and demonstrated in the unaveraged Hill-plus-nonspherical-potential model (Russell and Brinckerhoff 2009). Critical inclinations and frozen orbits about Moon were put emphasis when analyzing orbital characteristics of lunar artificial satellites (Carvalho et al. 2010). The critical inclination orbits for artificial lunar satellites were studied in the lunar potential model including the $J_2$ and $C_{22}$ terms and lunar rotation (Tzirti et al. 2009), and the subsequent work appended $J_3$ and the Earth's perturbation into consideration (Tzirti et al. 2010).

The stability of orbits around Europa was studied using analytical and numerical methods and a set of stable initial conditions were predicted (Scheeres et al. 2001). Long lifetime science orbits about Europa were designed by reducing the original 3-degree-of-freedom system twice via averaging (Paskowitz and Scheeres 2006). The complete description of families of frozen around Europa and their bifurcations were provided based on Lie transforms (San-Juan et al. 2006). Using two successive Lie-Deprit transforms, frozen orbits about Mercury in the averaged problem together with their stability were obtained, and the tests agreed well with the non-averaged model (Lara et al. 2010a). The locations of the stable and unstable Mercurial frozen orbits were determined using a Hamiltonian formalism, and the results were verified by numerical integrations of the complete systems (Delsate et al. 2010). By averaging the Hill problem up to six order based on Lie transforms, stable frozen orbits about Enceladus were predicted, and then checked in the non-averaged model (Lara et al. 2010b). Martian frozen orbits and critical inclination orbits were ever analyzed based on the mean element theory and the PSODE (particle swarm optimization combined with differential evolution) algorithm (Liu et al. 2010).

Among all the perturbations of the orbits around Mars, the major concern of this study is the Martian gravitational asphericity effect. It can be shown how the gravitational asphericity effect encourages and influences Martian frozen orbits independently. Other perturbations, including the three-body perturbation, atmospheric forces, and solar radiation pressure are not considered in the present paper. The work for Martian frozen orbits in the realistic ephemeris model is in progress.

Two different analytical methods are applied to analyze frozen orbits around Mars, and both are valid for quasi-circular orbits only. This paper is organized as follows. In Sect. 2, preliminary numerical investigations are employed to determine the number of zonal terms that should be considered in analytical investigations. The simulations show that zonal harmonics up to $J_9$ is enough to be involved in the analyses. In Sect. 3, an analytical method based on Lagrangian formulations is introduced to provide a portrait of families of Martian frozen orbits. In Sect. 4, analytical expansions based on Lie transforms are used to locate these orbits accurately. The contours of the Hamiltonian in the phase plane are also shown to illustrate the origins and evolutions of the equilibria. The two methods are compared with each other, and share similar results. Both discover four families of Martian frozen orbits: three families with small eccentricities and one family near the critical inclination. The analytical methods are examined in an $80 \times 80$ Mars gravity field, and the simulations show that the analytical methods can provide accurate enough results. In Sect. 5, the stability of families of Martian frozen orbits is estimated by calculating the trace of the monodromy matrix, and it shows that Martian frozen orbits over the range of inclinations $i \in [0, 90°]$ are approximately linearly stable. It should be noticed that the goal of this study is not to explore novel analytical theory, but to investigate characteristics of Martian frozen orbits, and search for proper initial conditions for numerical correction methods in the more complex models. It is hoped that the design principles of Martian frozen orbits in this paper will be helpful for future Mars exploration.

## 2 Impact of Different Terms of Zonal Harmonics

The model of Mars gravity field used in this study is the Goddard Mars Model 2B (GMM-2B), complete to degree and order 80 (Lemoine et al. 2001). For analytical investigations, it is impossible to consider all spherical harmonic coefficients because the algebraic manipulations in the $80\times80$ gravity field are too complicated. Therefore, it is necessary to determine how many terms should be involved in analytical investigations in order to make the process inexpensive but accurate enough.

The perturbations attributed to zonal harmonics are the most important factors to affect the motion of the spacecraft orbiting around Mars, while the perturbations attributed to tesserals can be neglected because they oscillate periodically (Smith 1986). To make clear the contribution of zonal harmonics, the simulations are conducted by increasing the number of zonal harmonics involved in the Mars gravity model, and the results are compared with the values when considering the $80\times80$ Mars gravity model GMM-2B. Thus, the contribution of zonal harmonics could be known from the plots. Figure 1 presents evolutions of the eccentricity and argument of pericentre for the same initial condition when considering different terms of spherical harmonic coefficients. If the calculations stop with $J_5$, the deviation of argument of pericentre from the value predicted in the $80\times80$ gravity model is about 40 degrees after five years; while if the calculations stop with $J_7$, deviations of the eccentricity and argument of pericentre are small; if the calculations stop with $J_9$, the deviations would be further reduced. Therefore, the Mars gravity model including zonal harmonics up to $J_9$ is enough to reach a good approximation.

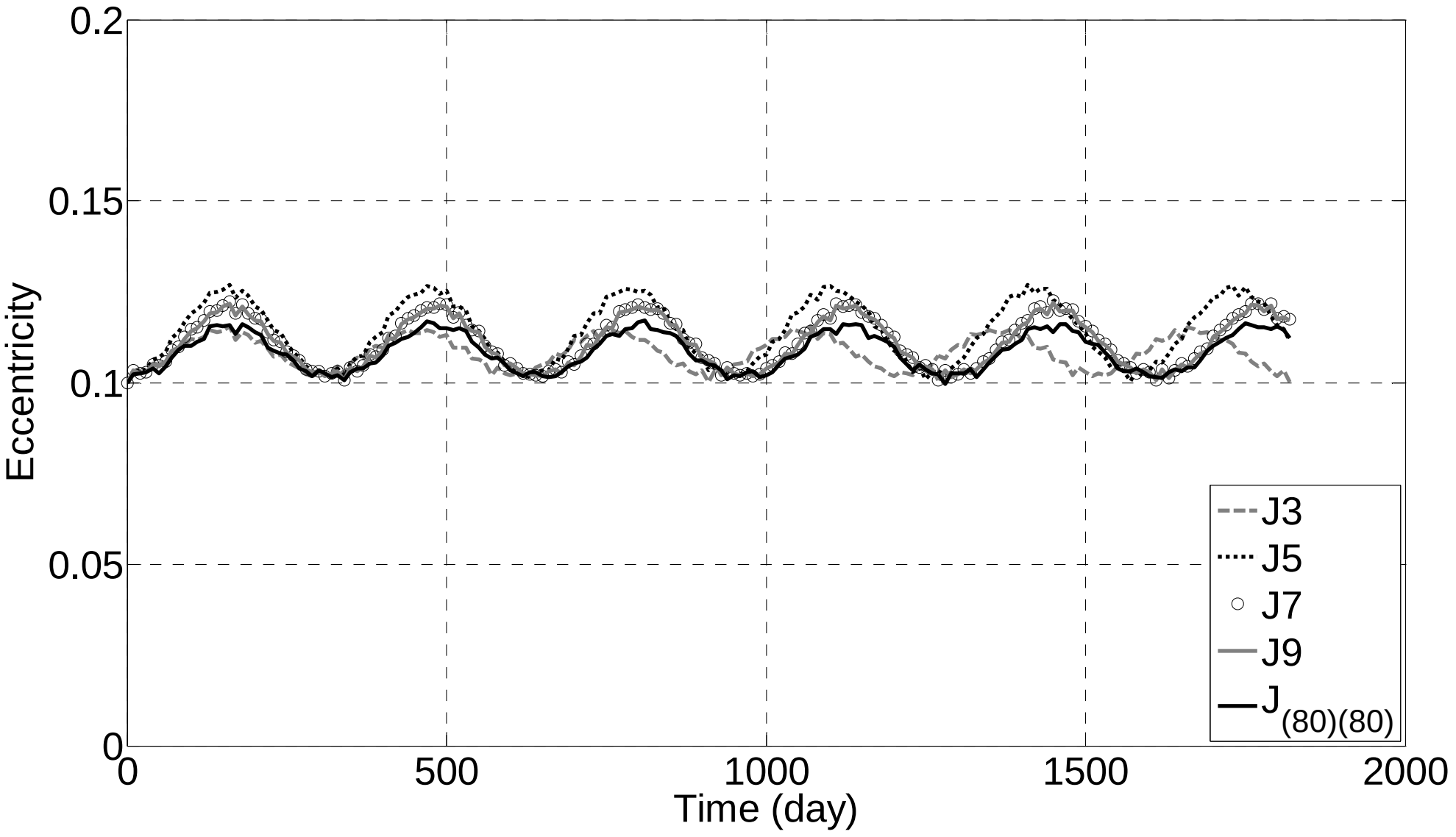

(**a**) Evolution of the eccentricity over one year.

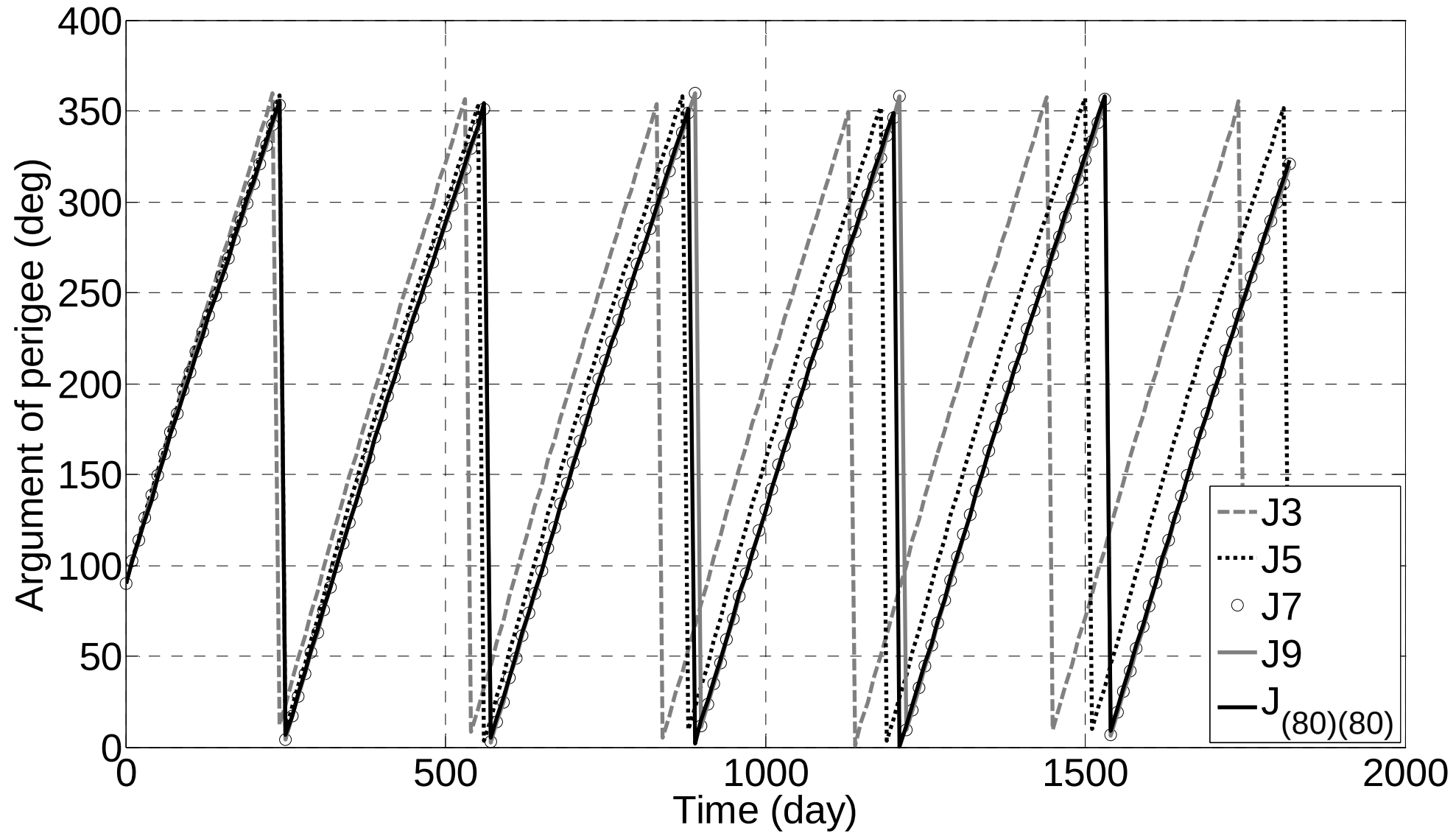

(**b**) Evolution of the argument of pericentre over one year.

**Fig. 1** Evolutions of the eccentricity and argument of pericentre over one year when considering different terms of spherical harmonic coefficients.

## 3 The Analytical Method Based on Lagrangian Formulations

The Mars disturbing geopotential is modeled using the expression (Kaula 1966)

$$U = \frac{\mu}{r}\left[\sum_{l=1}^{\infty} C_l \left(\frac{R_e}{r}\right)^l P_l\left(\sin\varphi\right) + \sum_{l=1}^{\infty}\sum_{m=1}^{l}\left(\frac{R_e}{r}\right)^l P_{lm}\left(\sin\varphi\right)\left(C_{lm}\cos m\lambda + S_{lm}\sin m\lambda\right)\right], \quad (1)$$

where $\mu$ is the Martian gravitational constant; $R_e$ is the reference radius of Mars; $r$ is the position of the spacecraft; $P_l$ is the Legendre function of degree $l$; $P_{lm}$ is the associated Legendre function of degree $l$ and order $m$; $C_l$ are the coefficients of zonals; $C_{lm}$ and $S_{lm}$ are the coefficients of the spherical harmonic expansion; $\varphi$ is the latitude of body-fixed coordinate system; and $\lambda$ is the longitude of the body-fixed coordinate system. Here, another symbol is introduced to represent the zonal harmonic coefficient

$$J_l = -C_l. \quad (2)$$

By transformation from spherical coordinates to Keplerian elements, the disturbing geopotential can be written as (Kaula 1966)

$$V = \sum_{l=1}^{\infty}\frac{\mu R_e^l}{r^{l+1}}\sum_{m=0}^{l}\sum_{p=0}^{l} F_{lmp}\left(i\right)\sum_{q=-\infty}^{\infty} G_{lpq}\left(e\right) S_{lmpq}\left(\omega, M, \Omega, \theta\right), \quad (3)$$

where

$$S_{lmpq} = \begin{bmatrix} C_{lm} \\ -S_{lm} \end{bmatrix}_{l-m\ odd}^{l-m\ even} \cos\left[(l-2p)\omega + (l-2p+q)M + m(\Omega-\theta)\right]$$
$$+\begin{bmatrix} S_{lm} \\ C_{lm} \end{bmatrix}_{l-m\ odd}^{l-m\ even} \sin\left[(l-2p)\omega + (l-2p+q)M + m(\Omega-\theta)\right]. \tag{4}$$

Here the variables $F_{lmp}(i)$ and $G_{lpq}(e)$ are functions of inclination and eccentricity, respectively; $e$ is eccentricity; $i$ is inclination; $\omega$ is argument of pericentre; $M$ is mean anomaly; $\Omega$ is right ascension of the ascending node; and $\theta$ is the sidereal time at the prime meridian.

The effect of tesseral harmonics are omitted, so that $m=0$. Further more, the disturbing potential can be averaged over the mean anomaly by eliminating the short-period terms, thus

$$l-2p+q=0. \tag{5}$$

Then, the disturbing potential only contains secular and long-period terms, and is rearranged as

$$V = -\sum_{l=1}^{\infty} \frac{\mu R_e^l}{r^{l+1}} \sum_{p=0}^{l} F_{l0p}(i) G_{lp(2p-l)}(e) J_l \begin{bmatrix} \cos\left[(l-2p)\omega\right] \\ \sin\left[(l-2p)\omega\right] \end{bmatrix}_{l\ odd}^{l\ even}. \tag{6}$$

By substituting Eq. (6) into the Lagrange's planetary equations, the long-period variation rates of the eccentricity and argument of pericentre arising from the odd degree zonal harmonics can be obtained, respectively (Rosborough and Ocampo 1991).

$$\dot{e}_l = 2n\frac{\left(1-e^2\right)^{1/2}}{e}\left[\sum_{k=1}^{\infty}\left(\frac{R_e}{r}\right)^{2k+1} F_{(2k+1)0k} G_{(2k+1)k(-1)} J_{2k+1}\right]\cos\omega, \tag{7}$$

$$\dot{\omega}_l = 2n\left\{\sum_{k=1}^{\infty}\left(\frac{R_e}{r}\right)^{2k+1}\left[\frac{\cos i}{\left(1-e^2\right)^{1/2}\sin i}F'_{(2k+1)0k}G_{(2k+1)k(-1)}\right.\right.$$
$$\left.\left.-\frac{\left(1-e^2\right)^{1/2}}{e}F_{(2k+1)0k}G'_{(2k+1)k(-1)}\right]J_{2k+1}\right\}\sin\omega\,, \tag{8}$$

where $n$ is the mean angular velocity, and the primes on the functions $F(i)$ and $G(e)$ indicate differentiation with respect to their respective variables.

Similarly, the secular variation rates of the eccentricity and argument of pericentre arising from the even degree zonal harmonics can be obtained, respectively (Rosborough and Ocampo 1991).

$$\dot{e}_s = 0\,, \tag{9}$$

$$\dot{\omega}_s = n\sum_{k=1}^{\infty}\left(\frac{R_e}{r}\right)^{2k}\left[\frac{\cos i}{\left(1-e^2\right)^{1/2}\sin i}F'_{(2k)0k}G_{(2k+1)k0}-\frac{\left(1-e^2\right)^{1/2}}{e}F_{(2k)0k}G'_{(2k)k0}\right]J_{2k}\,. \tag{10}$$

For frozen orbits, the average variation rates of the eccentricity and argument of pericentre are set equal to zero.

$$\dot{\bar{e}} = \dot{e}_l + \dot{e}_s = \dot{e}_l = 0\,, \tag{11}$$

$$\dot{\bar{\omega}} = \dot{\omega}_l + \dot{\omega}_s = 0\,. \tag{12}$$

Equation (11) can be satisfied by setting $\omega$ to be either 90 or 270 degrees. Turning to Eq. (12), the required eccentricity can be solved when setting the values of the semimajor axis and inclination.

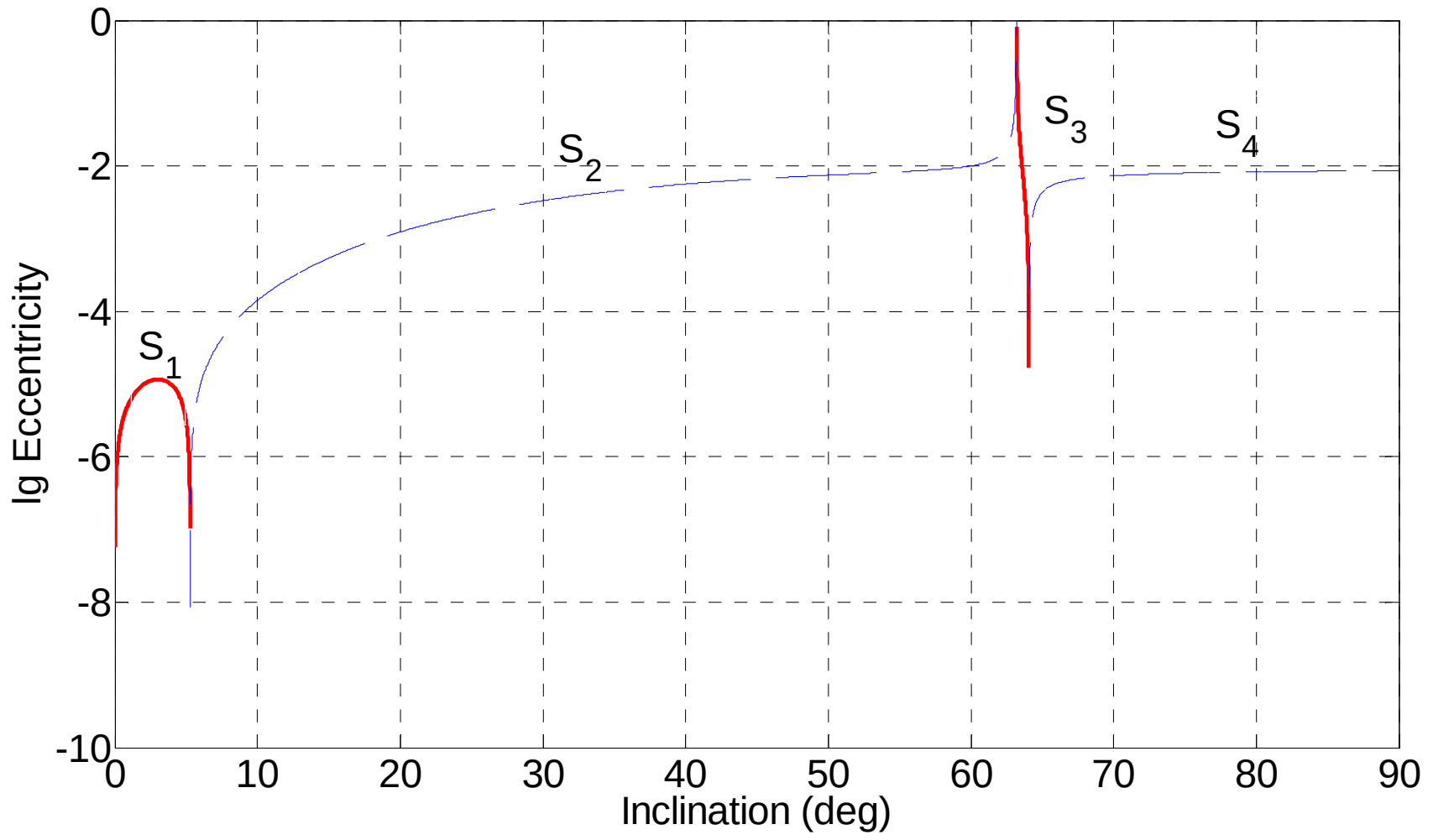

**Fig. 2** The logarithm of Martian frozen eccentricities vs. inclinations. The results are calculated based on Lagrangian formulations for the semimajor axis $a = 3597\ km$ in the zonal gravity field up to $J_9$. Solid lines in red correspond to $\omega = 90°$ and dashed lines in blue to $\omega = 270°$.

Figure 2 shows that Martian frozen orbits over the range of inclinations $i \in [0, 90°]$ can be divided into four families: the family $S_1$ far away from the critical inclination with small eccentricity and the argument of pericentre of argument equal to 90 degrees; the families $S_2$ and $S_4$ far away from the critical inclination with small eccentricity and the argument of pericentre equal to 270 degrees; the family $S_3$ close to the critical inclination with the argument of pericentre equal to 90 degrees. When comparing with Fig. 4 given by (Coffey et al. 1994) for Earth frozen orbits, it can be found that a new frozen family $S_1$ with very low eccentricity appears in Fig. 2, which is not present in Coffey et al's paper due to the difference between the Martian zonal harmonics and the Earth's; while the other three frozen families of the two figures are analogous. It is also noticed that at the boundary point of the families $S_1$ and $S_2$, there exist one frozen orbit with extremely small eccentricity, which is of special interest because it can provide much convenience for Mars reconnaissance.

It is evident that for the majority of Martian frozen orbits, argument of pericentre are kept at 270 degrees. That is because $J_3$ that is dominant among the odd degree zonals has the same sign with $J_2$. The effect of $J_3$ can be seen in the previous papers (Paskowitz and Scheeres 2006; Lara et al. 2010a; Delsate et al. 2010). While for a minority of ones with low inclination, argument of pericentre are possible to be kept at 90 degrees due to the effect of the higher degree odd zonals, which is an interesting phenomenon. In order to present details of Martian frozen orbits with low inclination, the required eccentricity contoured over the range of inclinations and semimajor axes ( $i \in [0, 20°]$ and $a \in [3397, 4397]\, km$ ) are given in Fig. 3. It shows that for a minority of Martian frozen orbits with low inclination and low altitude, argument of pericentre can be kept at 90 degrees. For the family $S_3$, argument of pericentre can also be kept at 90 degrees due to the singularity of the critical inclination.

In order to verify the behavior of Martian frozen orbits derived according to Lagrangian formulations, one of them is calculated in the $80 \times 80$ gravity field over one year. The set of orbital elements at the initial time are $a_0 = 3597\, km$ , $i_0 = 50°$ , $e_0 = 0.00746298$, and $\omega_0 = 270°$ . The evolutions of the eccentricity and argument of pericentre are shown by the radial graph (where $e$ is radial and $\omega$ is counterclockwise) in Fig. 4, and the result achieves a good property as an approximately frozen orbit. The eccentricity oscillates around $e_0$, and the amplitude is about 0.0025; the argument of pericentre oscillates around $\omega_0$, and the amplitude is about 20 degrees.

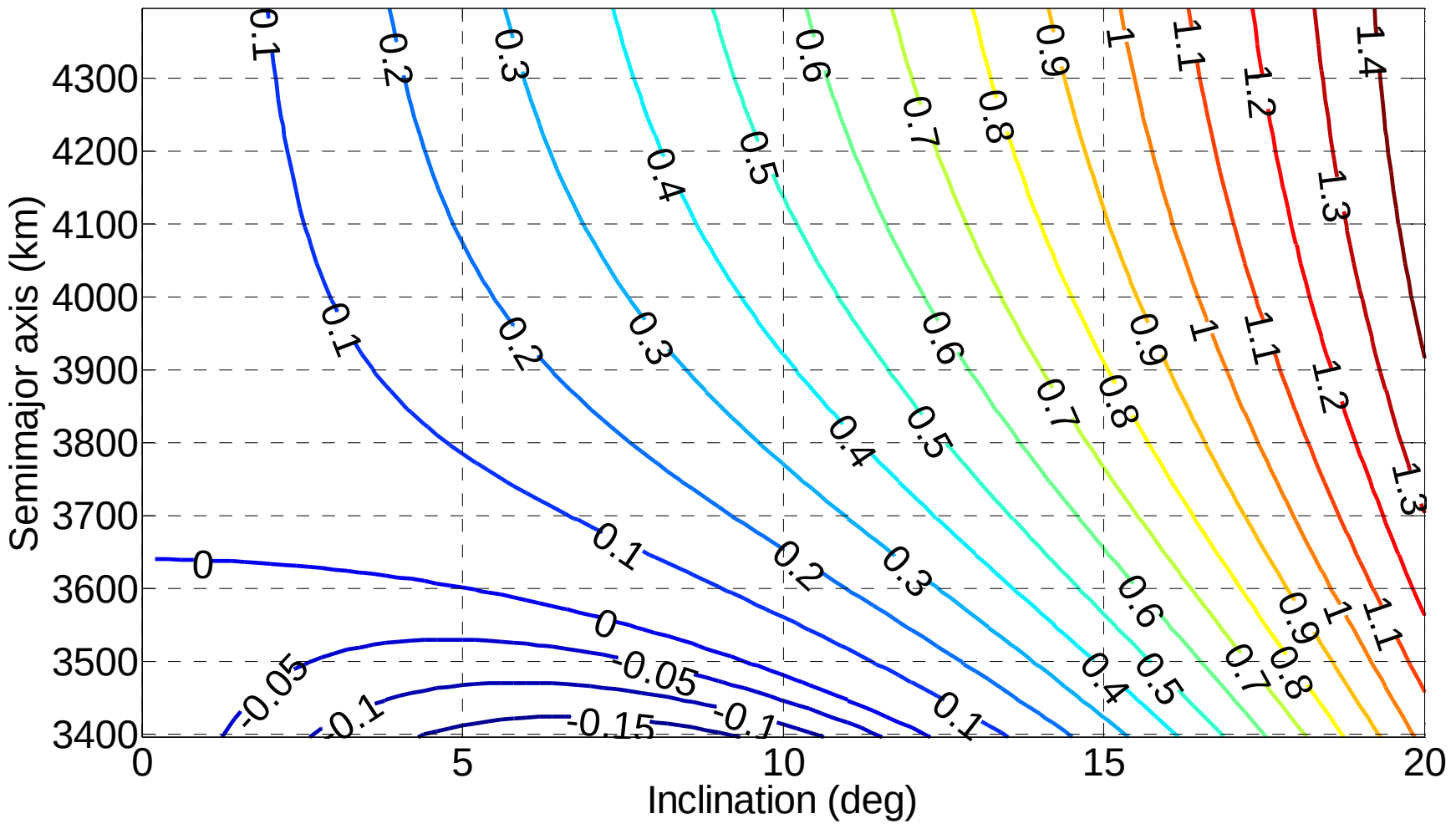

**Fig. 3** The contours of the frozen eccentricity over a wide range of inclination and semimajor axis ($i \in [0, 20°]$ and $a \in [3397, 4397]\ km$) for the zonal gravity field up to $J_9$. The values of the eccentricity are magnified 100 times.

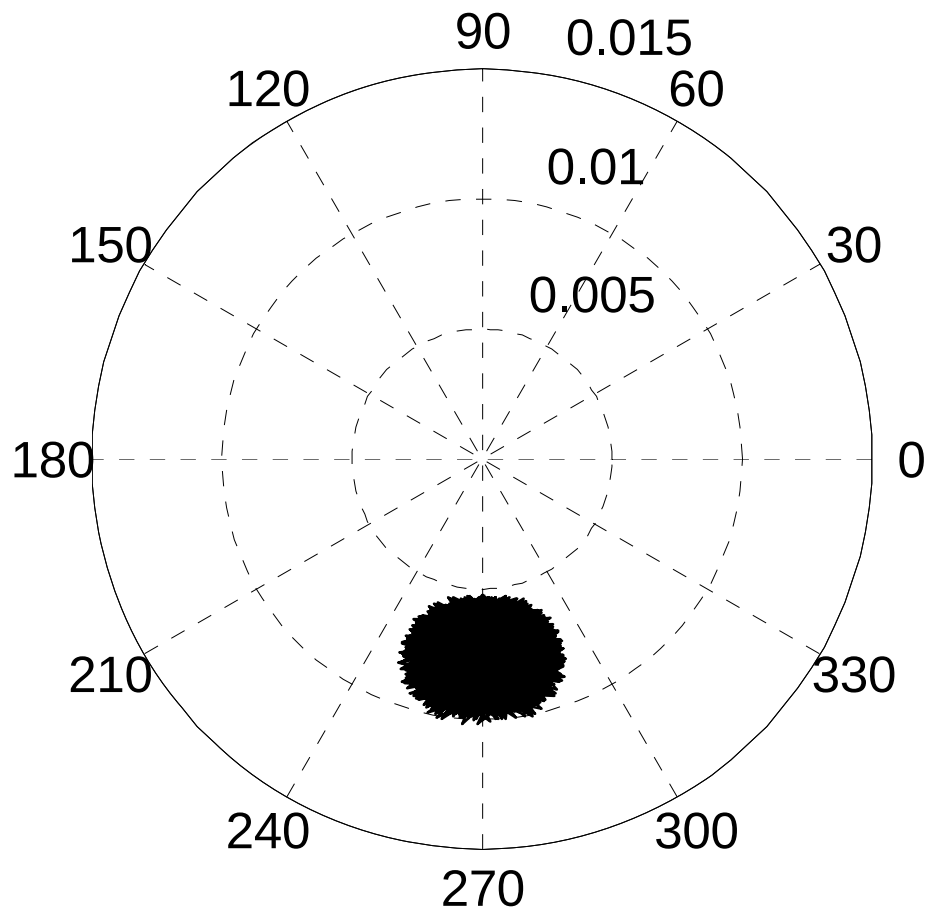

**Fig. 4** $e-\omega$ evolution over one year in the $80\times 80$ gravity field for $a_0 = 3597\,km$ and $i_0 = 50°$. $e$ is radial and $\omega$ is counter-clockwise. The initial condition is calculated based on Lagrangian formulations.

## 4 Analytical Expansions Based on Lie transforms

Based on Lie transforms to the second order, the gravitational perturbation problem can be averaged over the mean anomaly to obtain the normalized Hamiltonian, which takes the form of a Fourier series of argument of pericentre with coefficients as algebraic functions of eccentricity. The resulting Hamiltonian up to $J_9$ used in this section was derived by Coffey et al. (1994) in Appendix, and the symbolic coefficients are replaced by the corresponding numbers from Mars potential.

The conjugate Delaunay variables used in this paper are written as

$$\begin{gathered} l = M,\ g = \omega,\ h = \Omega, \\ L = \sqrt{\mu a},\ G = \sqrt{\mu a\left(1-e^2\right)},\ H = \sqrt{\mu a\left(1-e^2\right)}\cos i, \end{gathered} \tag{13}$$

where $H$ is the polar component of the angular momentum. Expressed in Delaunay variables, the canonical equations can be obtained from the Hamiltonian.

$$\dot{g} = \frac{\partial \mathcal{H}}{\partial G},\ \dot{G} = -\frac{\partial \mathcal{H}}{\partial g}. \tag{14}$$

The system composed of the spacecraft and Mars is conservative, so one orbit around Mars corresponds to one value of the Hamiltonian. Thus, the contours of the Hamiltonian in the phase plane of $g$ and $G$ can show the evolutions of orbits, and the equilibria in the phase plane represent the frozen locations. When the value of $H$ passes through the locations of bifurcation, a global change would occur in the behavior of the contours of the Hamiltonian.

In order to make clear the origin of Martian frozen orbits, primarily the simplest case is considered, where only the representative of the even zonal harmonic $J_2$ and the representative of the odd zonal harmonic $J_3$ are involved. Then Eq. (14) can be further rearranged as

$$\begin{aligned}\dot{g} &= \frac{G}{p^4} R_e^2 \eta^3 \left( -\frac{15}{4} s^2 + 3 \right) \\ &+ J_2 \frac{R_e^4 G}{p^6} \left\{ \eta^5 \left( \frac{135}{128} s^4 + \frac{27}{32} s^2 - \frac{21}{16} \right) + \eta^4 \left( \frac{135}{16} s^4 - \frac{99}{8} s^2 + \frac{9}{2} \right) + \eta^3 \left( \frac{1155}{128} s^4 - \frac{645}{32} s^2 + \frac{165}{16} \right) \right. \\ &\left. + \left[ e^2 \eta^3 \left( \frac{495}{64} s^4 - \frac{279}{32} s^2 + \frac{21}{16} \right) + \eta^5 \left( \frac{45}{32} s^4 - \frac{21}{16} s^2 \right) \right] \cos 2g \right\} \\ &+ \frac{J_3}{J_2} \frac{R_e^3 G}{p^5} \left[ e\eta^3 \left( -15 s^3 + 9s \right) - \frac{\eta^5}{e} \left( \frac{15}{8} s^3 - \frac{3}{2} s \right) + \frac{e\eta^3}{s} \left( \frac{45}{8} s^2 - \frac{3}{2} \right) \right] \sin g + o\left( J_2^2 \right),\end{aligned} \tag{15}$$

$$\begin{aligned}\dot{G} = J_2 \frac{G^2}{p^6} \Bigg[ & R_e^4 \frac{J_4}{J_2^2} e^2 \eta^3 \left( -\frac{105}{32} s^4 + \frac{45}{16} s^2 \right) \sin 2g - R_e^3 \frac{J_3}{J_2^2} p e \eta^3 \left( \frac{15}{8} s^3 - \frac{3}{2} s \right) \cos g \\ & + R_e^4 e^2 \eta^3 \left( -\frac{45}{32} s^4 + \frac{21}{16} s^2 \right) \sin 2g \Bigg] + o\left( J_2^2 \right),\end{aligned} \tag{16}$$

where $p$ is semiparameter; $\eta = \sqrt{1-e^2}$ ; and $s = \sin i$ .

It is evident that Eq. (16) can be set equal to zero by choosing $g$ to be either 90 degrees or 270 degrees, which is in accordance with the results obtained according to Lagrangian formulations. Corresponding to $g = 90°$ and $g = 270°$ respectively, two different equations can be derived from Eq. (15), from which two bifurcations arise in the phase plane. When $g = 90°$ , the location for the first bifurcation can be obtained, and is denoted as $H_1$. The variable $H_1$ is the real root of the following equation.

$$\frac{5H_1^2}{L^2} - 1 + \frac{J_2}{5}\left(\frac{R_e}{a}\right)^2 + \frac{2\sqrt{5}J_3}{5J_2}\left(\frac{R_e}{a}\right)\sqrt{1 - \frac{5H_1^2}{L^2}} = 0 . \tag{17}$$

When $g = 270°$, the location for the second bifurcation denoted as $H_2$ is the real root of the following equation.

$$\frac{5H_2^2}{L^2} - 1 + \frac{J_2}{5}\left(\frac{R_e}{a}\right)^2 - \frac{2\sqrt{5}J_3}{5J_2}\left(\frac{R_e}{a}\right)\sqrt{1 - \frac{5H_2^2}{L^2}} = 0 . \tag{18}$$

The values of $H_1$ and $H_2$ can be obtained by numerical methods. Figure 5a presents the dependence of the locations for bifurcations on the semimajor axis. It can be seen the two curves are indistinguishable. As shown in Fig. 5b, the differences between the values of $H_1/L$ and $H_2/L$ for the same semimajor axis are very small, the order of magnitude of which is about $10^{-4}$.

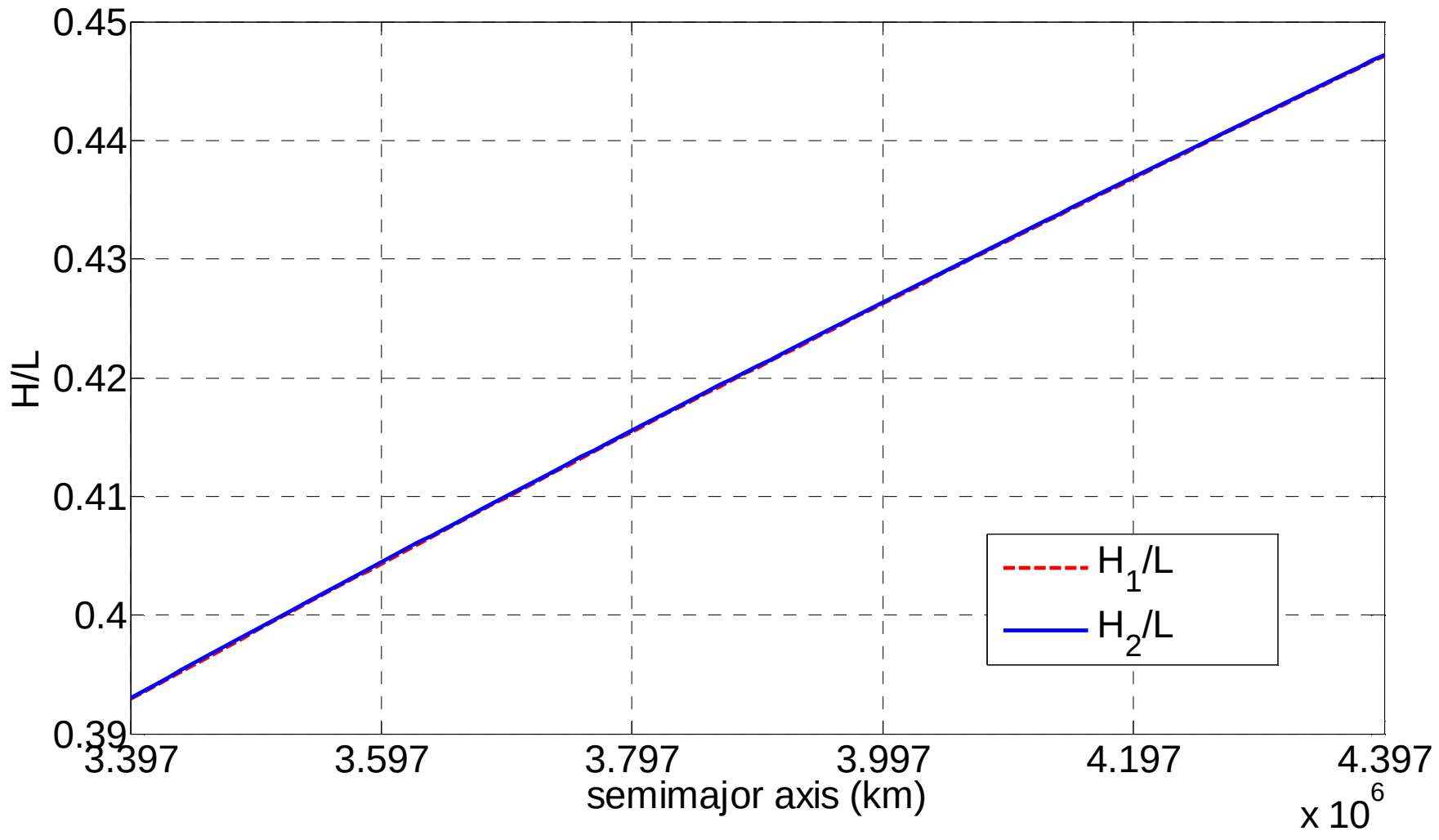

(**a**) $H_1$ /L and $H_2$ /L vs. semimajor axes.

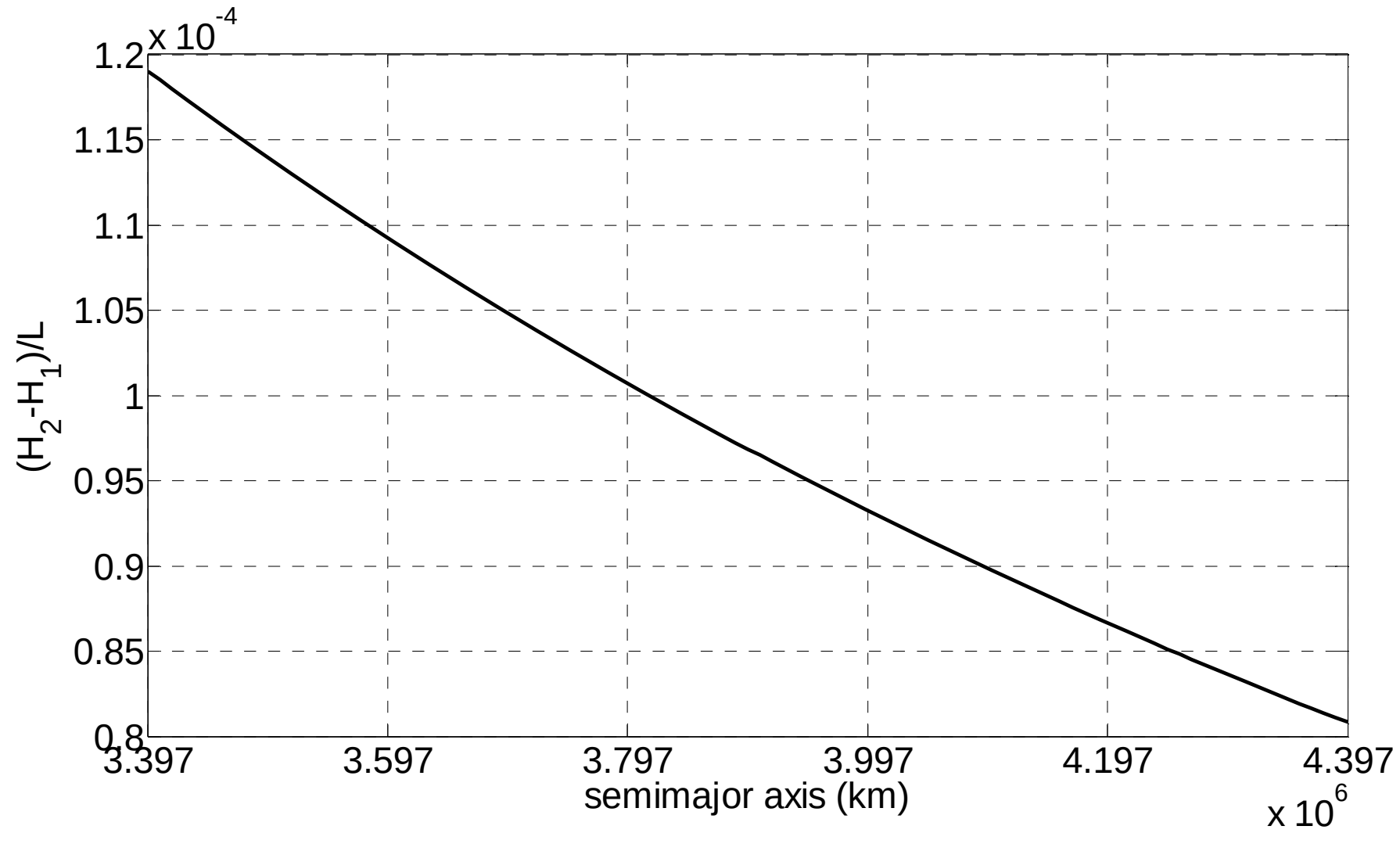

(**b**) Differences between $H_1/L$ and $H_2/L$ vs. semimajor axes.

Fig. 5 The locations for bifurcation as functions of the semimajor axis in the $J_2$+$J_3$ gravity field.

Figure 6 presents that with the variations of $H$ for the same semimajor axis $a = 3597\,\text{km}$ in the $J_2$+$J_3$ gravity field, two bifurcations occur in the phase plane. At $H = H_2$, a pitchfork bifurcation occurs from a central point to two central points and a saddle point; while at $H = H_1$, a saddle-node bifurcation gives rise to a saddle point and a central point. When considering zonal harmonics up to $J_9$, the evolution of the phase plane for $a = 3597\,\text{km}$ is a little different, which can be shown in Fig. 7. For $H = 0.7L$, there is only one central point in the phase plane of $g$ and $G$. Between the values of $H = 0.7L$ and $H = 0.4492L$, a saddle-node bifurcation gives rise to a saddle point and a central points. Then before the value of $H$ reaches $0.4486L$, a pitchfork bifurcation occurs from a saddle point to a central point and two saddle points. The phase plane for $H = 0$ represents the polar orbit, which is widely applicable to Martian missions, only has a saddle point at about $\omega = 270°$. Comparing with Fig. 9 given by (Coffey et al. 1994), it can be found that the evolution of the phase place with the variations of $H$ for Mars is more complex than that for Earth due to the characteristics of the Martian gravity field.

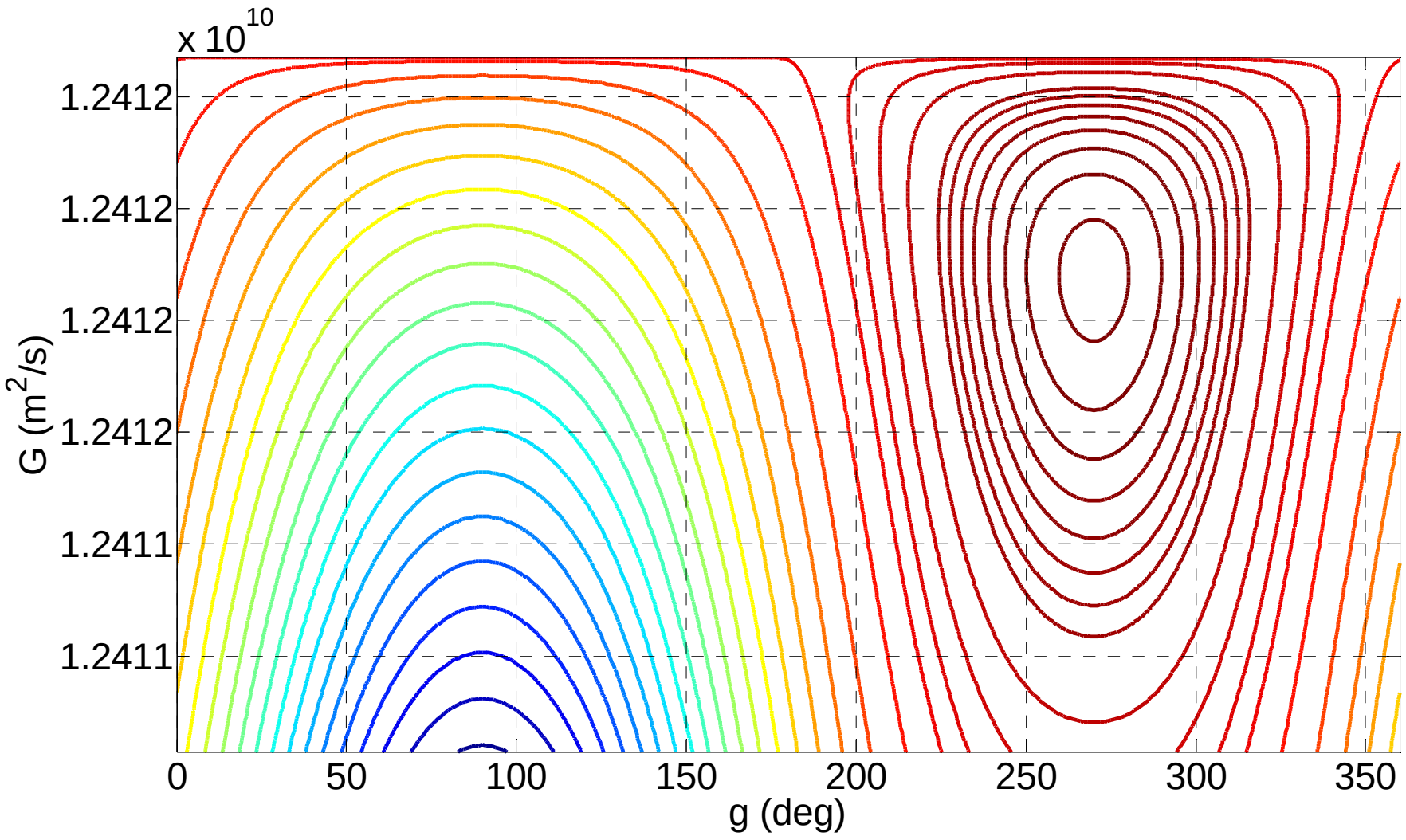

**(a)** $H > H_2$.

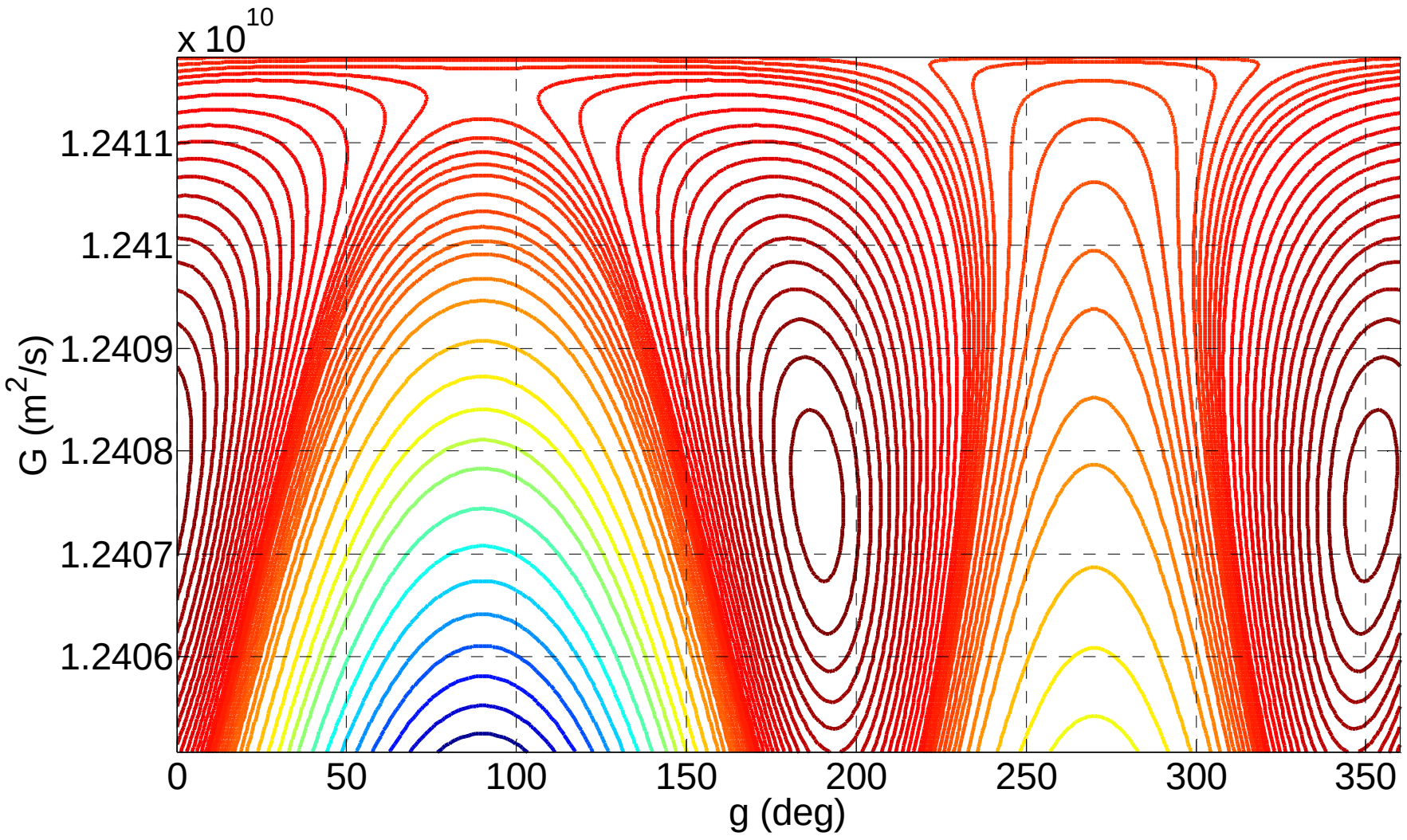

**(b)** $H_1 < H < H_2$.

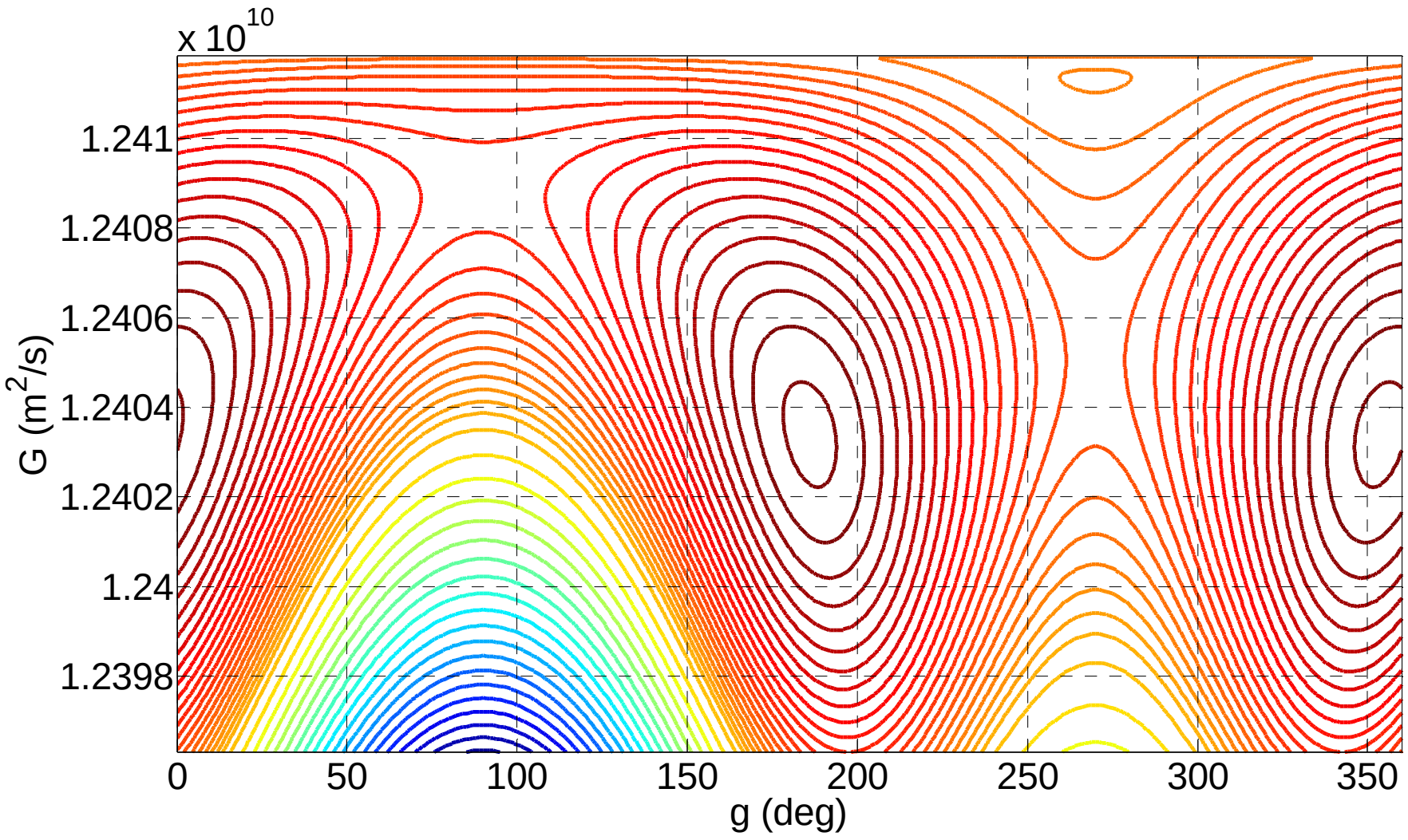

(**c**) $H<H_1$.

**Fig. 6** The contours of the Hamiltonian for $a = 3597\,\text{km}$ in the phase plane of $g$ and $G$ with different values of $H$ in the $J_2$+$J_3$ gravity field.

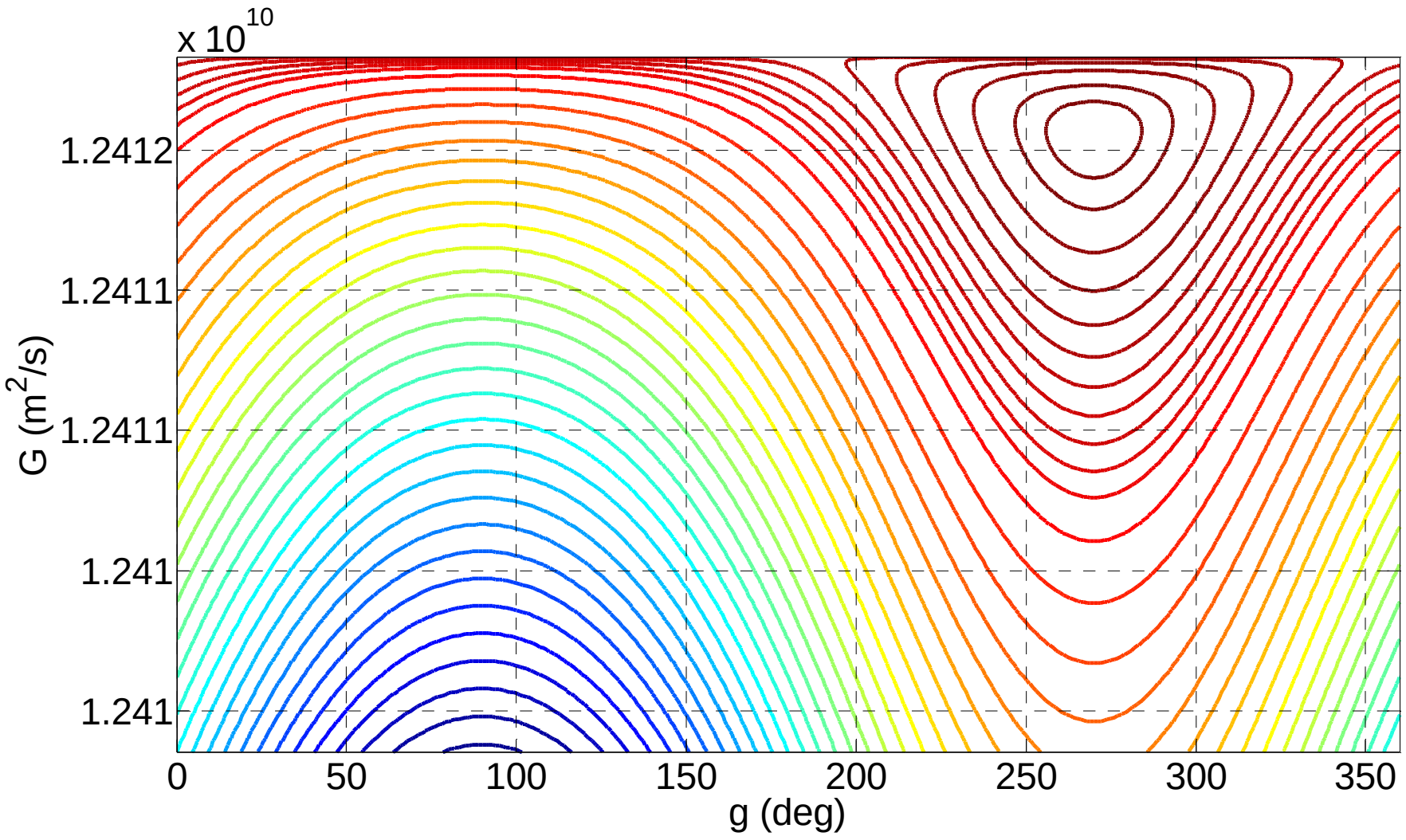

(**a**) $H$=0.7$L$.

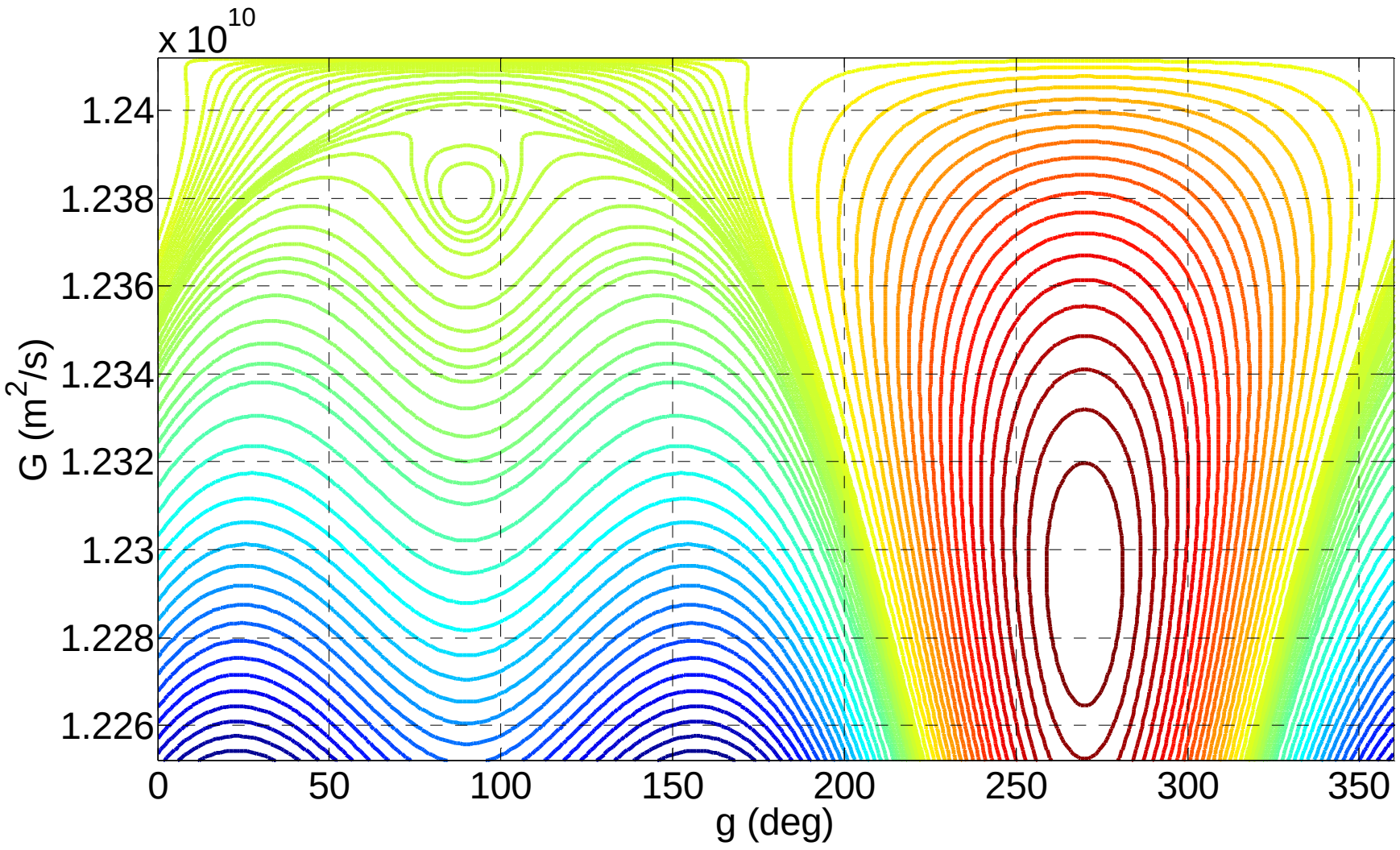

(**b**) $H$=0.4492$L$.

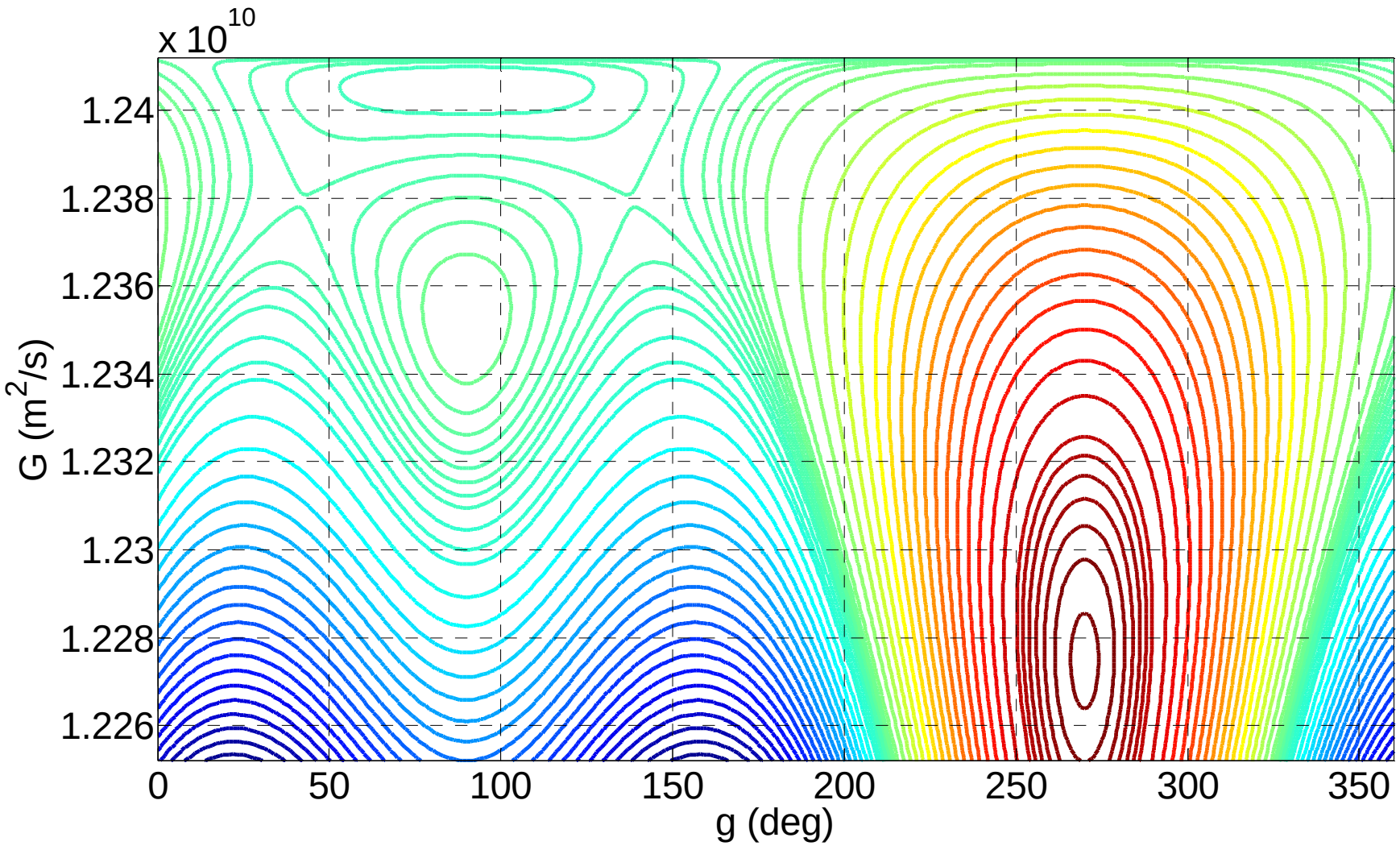

(**c**) $H$=0.4486$L$.

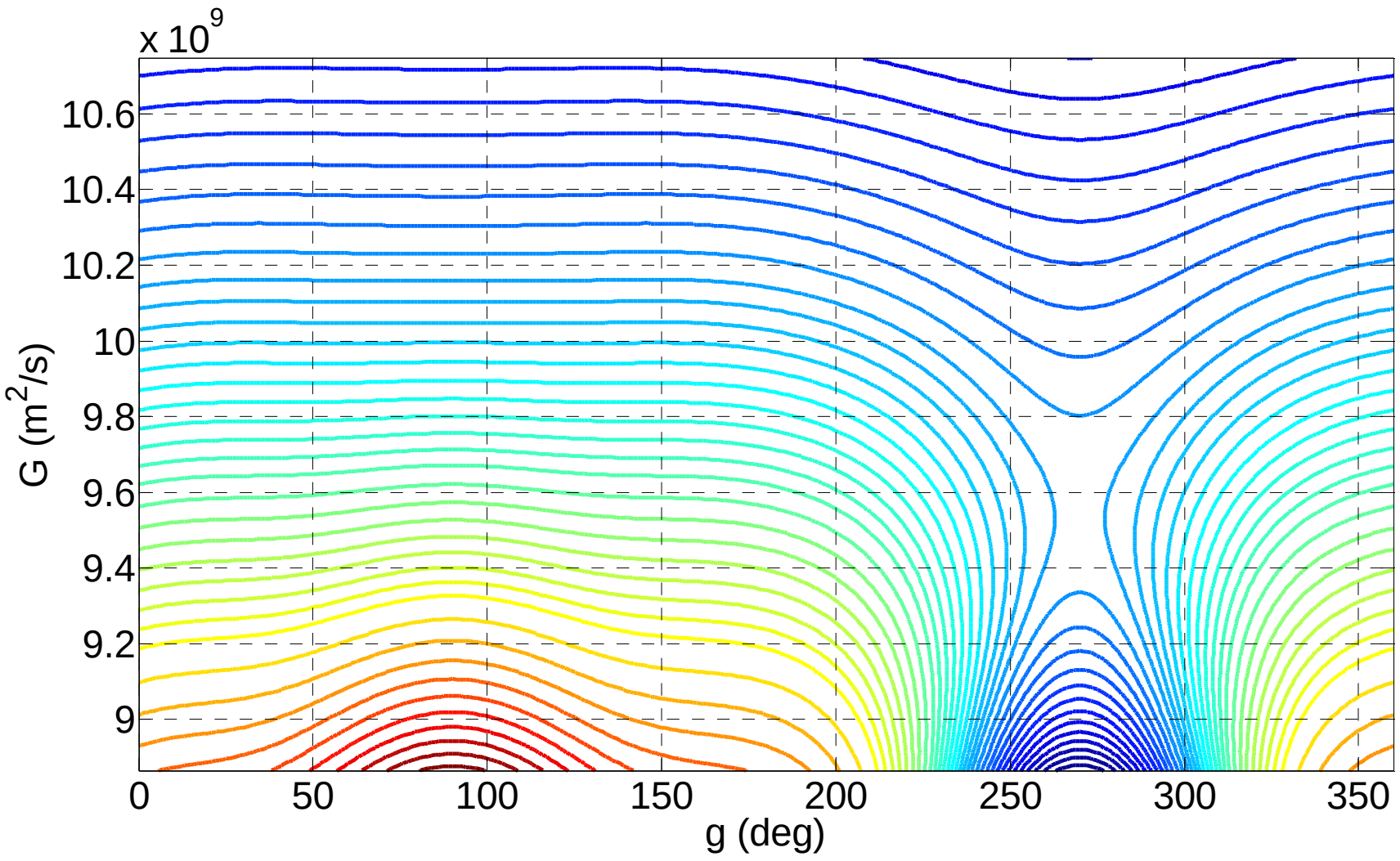

(**d**) *H*=0.

**Fig. 7** The contours of the Hamiltonian for $a = 3597\,\text{km}$ with different values of $H$ in the zonal gravity field up to $J_9$.

Figure 8 shows the dependence of the phase plane on the semimajor axis in the zonal gravity field up to $J_9$. For $a = 4397\,\text{km}$, only one central point exists in the phase plane. Between the values of $a = 4397\,\text{km}$ and $a = 3779\,\text{km}$, two central points arise. Then before the value of semimajor axis reaches $a = 3397\,\text{km}$ (which can be seen in Fig. 7(c)), a pitchfork bifurcation occurs from a central point to three equilibria, a central point and two saddle points. When the value of the semimajor axis is large enough, there is only one central point in the phase plane, which is different from Fig. 8 given by (Coffey et al. 1994).

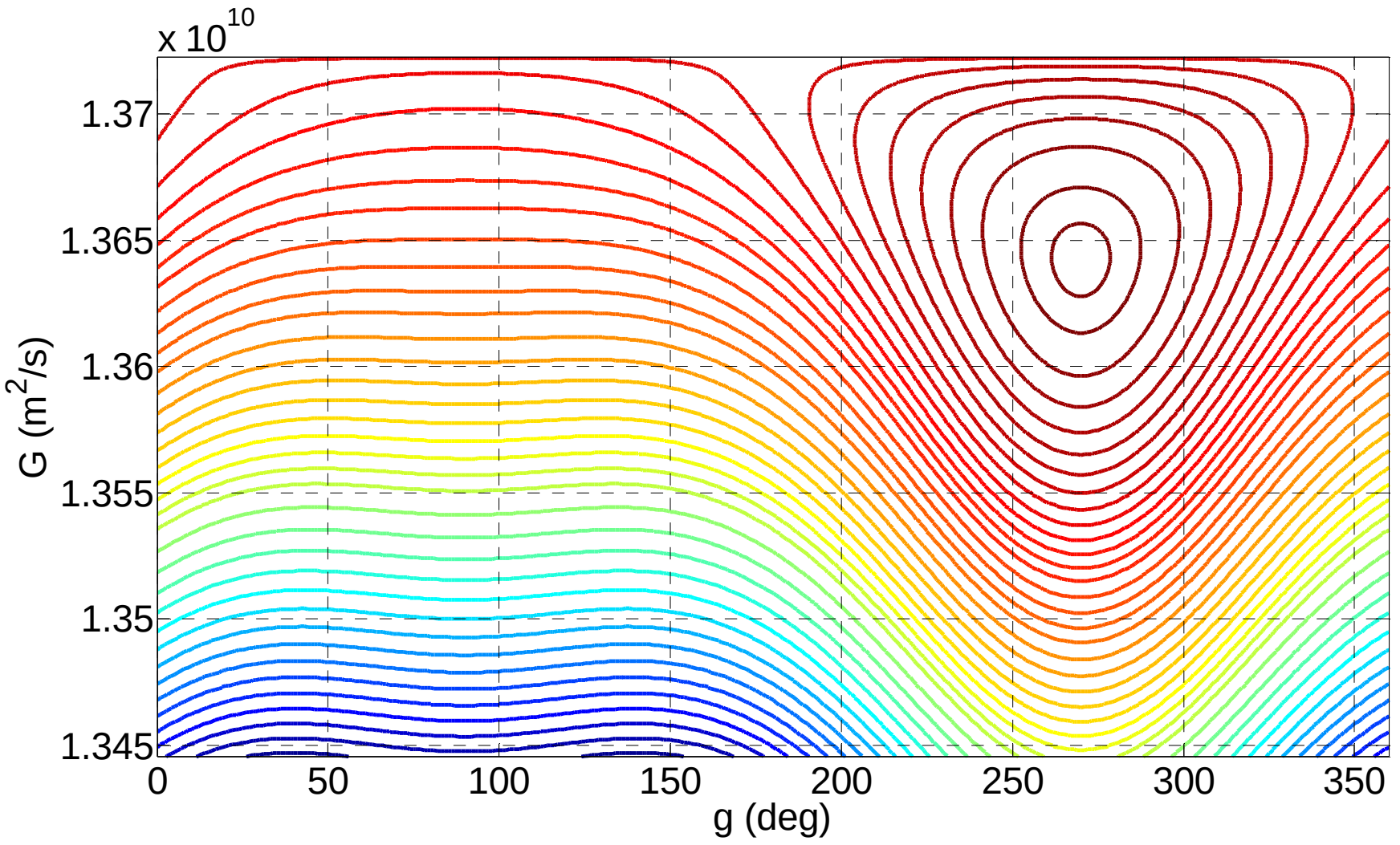

**(a)** $a = 4397\,\mathrm{km}$

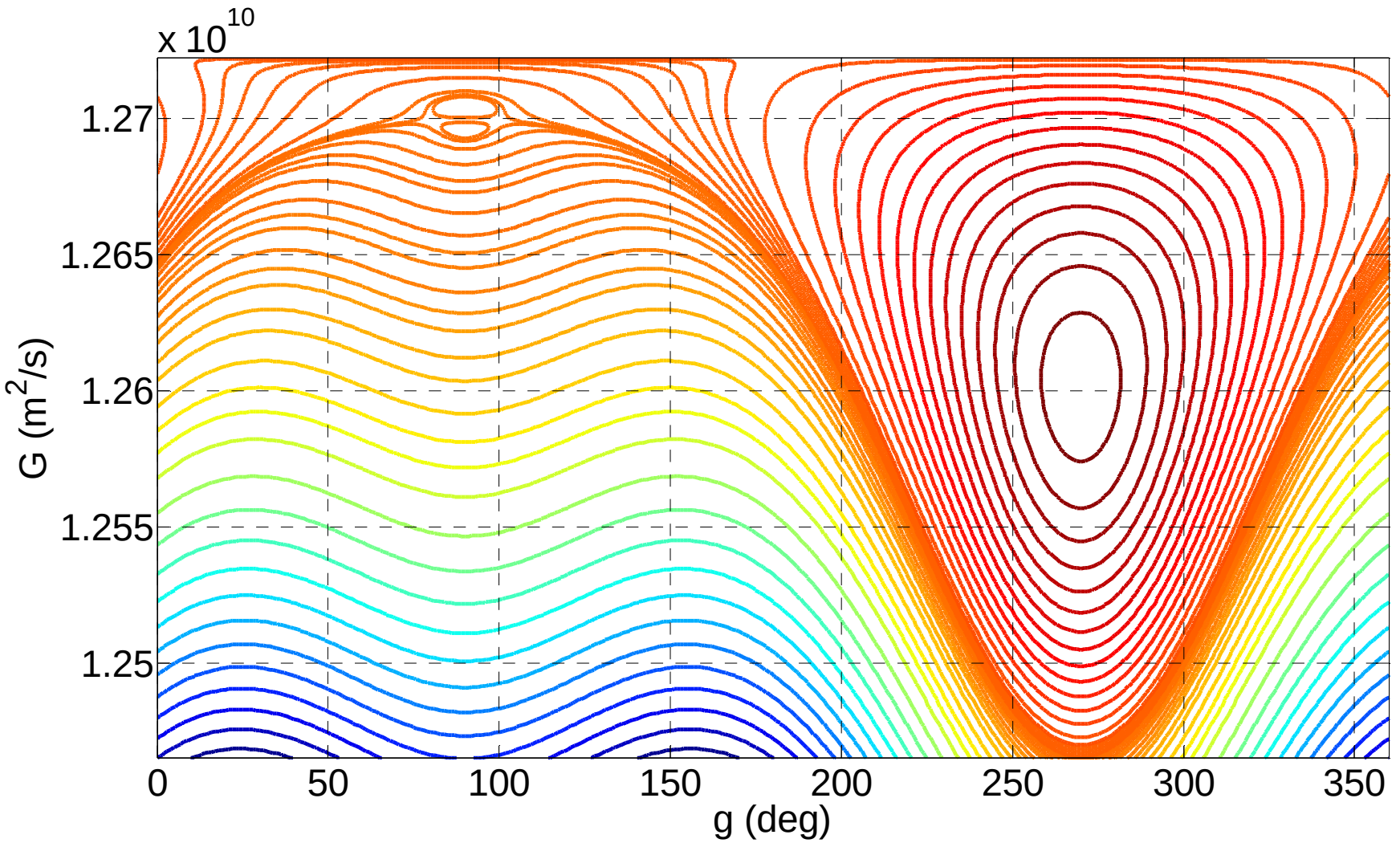

**(b)** $a = 3779\,\mathrm{km}$

**Fig. 8** The contours of the Hamiltonian for $H = 0.4486L$ in the phase plane of $g$ and $G$ with different values of semimajor axes in the zonal gravity field up to $J_9$.

The above illustrations are geometrical; however, they are not convenient to provide quantitative information. Thus, in the following, the analytical method for solving

Martian frozen orbits is given.

The differential equations of $g$ and $G$ are written in the form

$$\begin{aligned} \dot{g} &= f\left(g, G\right) = f_0 + J_2 f_1, \\ \dot{G} &= F\left(g, G\right) = F_0 + J_2 F_1, \end{aligned} \tag{19}$$

and it can be proved that

$$f_0 = \frac{3G}{4p^2}\left(\frac{R_e}{p}\right)^2 \eta^3\left(4 - 5s^2\right). \tag{20}$$

It is evident the equilibria of the system (19) exist in the vicinity of zero eccentricity ($e = 0$) or the critical inclination ($5s^2 = 4$). The value of the critical inclination can be easily obtained as $63.43494882°$. The solution of $f_1 = 0$ is complex, and in this case, the only perturbation of the motion is the long-period terms involving $J_2$, which is not considered in this paper.

For frozen orbits near the critical inclination, $s$ can be expanded at $s_0 = \sqrt{4/5}$ as

$$s = s_0 - J_2 f_1\left(s_0\right)\left[\left.\frac{\partial f_0}{\partial s}\right|_{s=s_0}\right]^{-1}, \tag{21}$$

where

$$\left.\frac{\partial f_0}{\partial s}\right|_{s=s_0} = -3\sqrt{5}\,\frac{G}{p^2}\left(\frac{R_e}{p}\right)^2 \eta^3, \tag{22}$$

$$f_1\big|_{s=s_0} = \frac{G}{p^2}\left\{\left(\frac{R_e}{p}\right)^4\left(-\frac{21}{40}\eta^5+\frac{27}{40}\eta^3\right)+\frac{\sqrt{5}J_3}{2eJ_2^2}\left(\frac{R_e}{p}\right)^3\left(-\frac{3}{5}\eta^5+\frac{3}{5}\eta^3\right)\sin g\right.$$
$$+\frac{J_4}{J_2^2}\left(\frac{R_e}{p}\right)^4\left(-\frac{15}{8}\eta^5+\frac{123}{40}\eta^3\right)$$
$$+\frac{\sqrt{5}J_5}{2eJ_2^2}\left(\frac{R_e}{p}\right)^5\left(\frac{573}{200}\eta^7-\frac{519}{50}\eta^5+\frac{63}{8}\eta^3\right)\sin g$$
$$+\frac{J_6}{J_2^2}\left(\frac{R_e}{p}\right)^6\left(\frac{13167}{3200}\eta^7-\frac{6531}{320}\eta^5+\frac{62559}{3200}\eta^3\right)$$
$$+\frac{\sqrt{5}J_7}{2eJ_2^2}\left(\frac{R_e}{p}\right)^7\left(-\frac{117663}{32000}\eta^9+\frac{902937}{32000}\eta^7-\frac{1767717}{3200}\eta^5+\frac{997227}{32000}\eta^3\right)\sin g$$
$$+\frac{J_8}{J_2^2}\left(\frac{R_e}{p}\right)^8\left(-\frac{371763}{128000}\eta^9+\frac{732753}{25600}\eta^7-\frac{8801793}{128000}\eta^5+\frac{1180179}{25600}\eta^3\right)$$
$$+\frac{\sqrt{5}J_9}{2eJ_2^2}\left(\frac{R_e}{p}\right)^9\left(\frac{307755}{640000}\eta^{11}-\frac{1193661}{160000}\eta^9+\frac{9821889}{320000}\eta^7-\frac{1450449}{32000}\eta^5\right.$$
$$\left.+\frac{13936923}{640000}\eta^3\right)\sin g. \tag{23}$$

While for frozen orbits with nearly zero eccentricity, $G$ can be expanded at $G = G_0$ as

$$G = G_0 - J_2 f_1(G_0)\left[\frac{\partial f_0}{\partial G}\bigg|_{G=G_0}\right]^{-1}. \tag{24}$$

where $G = G_0$ is equivalent to $e = 0$. In order to prevent the divisor $e$ appearing in the denominator, both sides of Eq. (24) are multiplied by $e$.

$$(eG) = (eG_0) - J_2\left[ef_1(G_0)\right]\left[\frac{\partial f_0}{\partial G}\bigg|_{G=G_0}\right]^{-1}, \tag{25}$$

where

$$\frac{\partial f_0}{\partial G}\bigg|_{e=0} = \frac{3}{4p^2}\left(\frac{R_e}{p}\right)^2\left(30s^2-26\right), \tag{26}$$

$$
\begin{aligned}
\left.(ef_1)\right|_{e=0} = \frac{G}{p^2}\Bigg[ & \frac{J_3}{J_2^2}\left(\frac{R_e}{p}\right)^3\left(-\frac{15}{8}s^3+\frac{3}{2}s\right)+\frac{J_5}{J_2^2}\left(\frac{R_e}{p}\right)^5\left(-\frac{315}{32}s^5+\frac{105}{8}s^3-\frac{15}{4}s\right) \\
& +\frac{J_7}{J_2^2}\left(\frac{R_e}{p}\right)^7\left(-\frac{45045}{1024}s^7+\frac{10395}{128}s^5-\frac{2835}{64}s^3+\frac{105}{16}s\right) \\
& +\frac{J_9}{J_2^2}\left(\frac{R_e}{p}\right)^9\left(-\frac{765765}{4096}s^9+\frac{225225}{512}s^7-\frac{45045}{128}s^5+\frac{3465}{32}s^3-\frac{315}{32}s\right)\Bigg]\sin g.
\end{aligned}
\tag{27}
$$

According to Eq. (21), for frozen orbits near the critical inclination, the inclination can be determined when given the value of the eccentricity; while according to Eq. (24), for orbits far away the critical inclination, the frozen eccentricity can be calculated if setting the value of the inclination. Therefore, Martian frozen orbits at all inclinations can be derived, as shown in Fig. 9. Similarly to the conclusion obtained from Lagrangian formulations, four families of Martian frozen orbits are discovered: the family $S_1'$ far away from the critical inclination with small eccentricity and the argument of pericentre of argument equal to 90 degrees; the families $S_2'$ and $S_4'$ far away from the critical inclination with small eccentricity and the argument of pericentre equal to 270 degrees; the family $S_3'$ near the critical inclination with the argument of pericentre equal to 90 degrees. There also exists one frozen orbit with extremely small eccentricity at the boundary point of the families $S_1'$ and $S_2'$.

Figure 10 shows the differences in frozen eccentricities of the two different methods (Lagrangian formulations and Lie transforms) over the range of inclinations $i \in [0, 90°]$. It can be seen that for orbits far away from the critical inclination, the results of the two different methods are well consistent with each other, and the differences in frozen eccentricities are about 1–6 orders of magnitude lower than frozen eccentricities themselves. While for orbits close to the critical inclination, the differences in frozen eccentricities of the two different methods are large due to the singularity of the critical inclination.

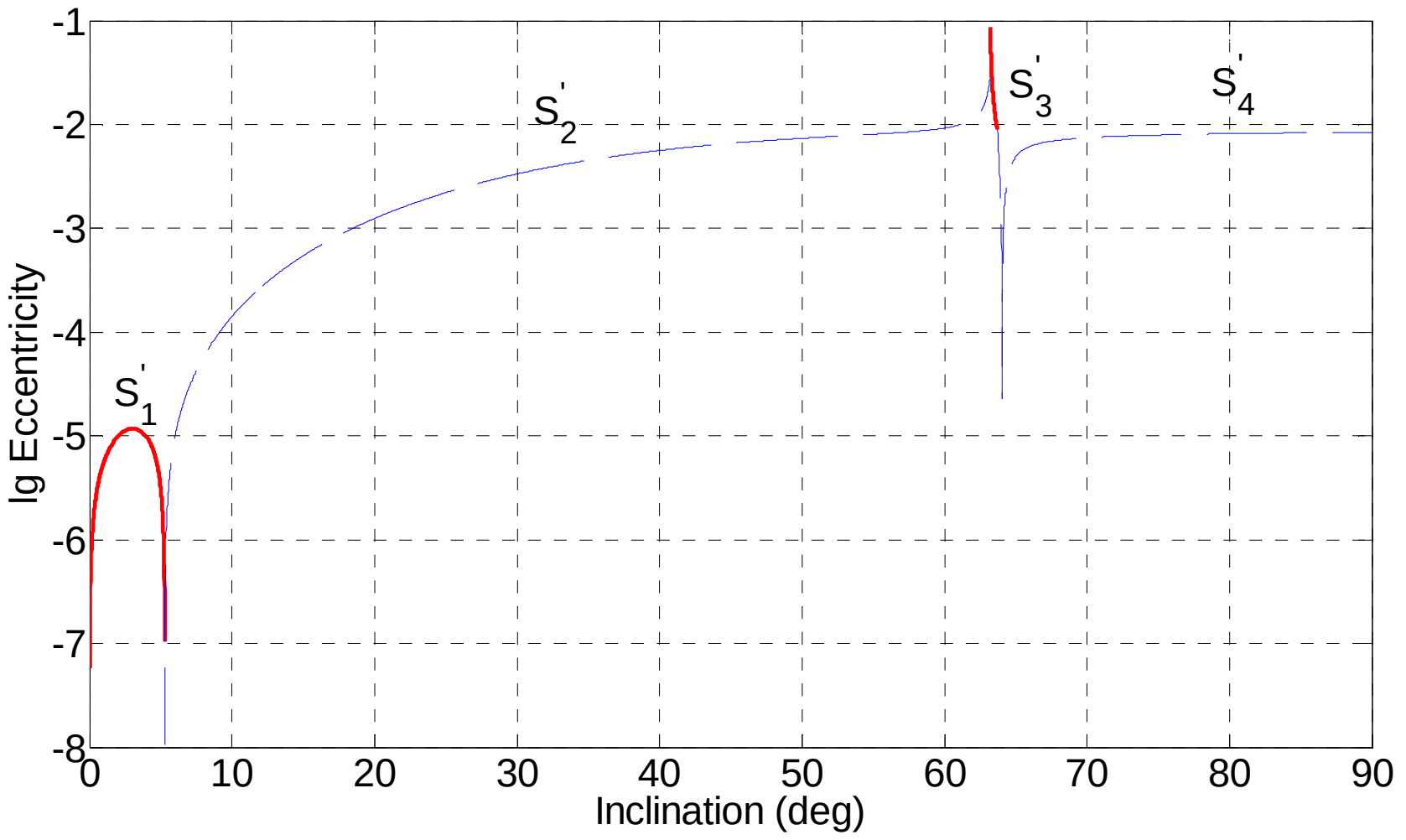

**Fig. 9** The logarithm of Martian frozen eccentricities vs. inclinations. The results are calculated based on Lie transforms for $a = 3597\ km$ in the zonal gravity field up to $J_9$. Solid lines in red correspond to $\omega = 90°$ and dashed lines in blue to $\omega = 270°$.

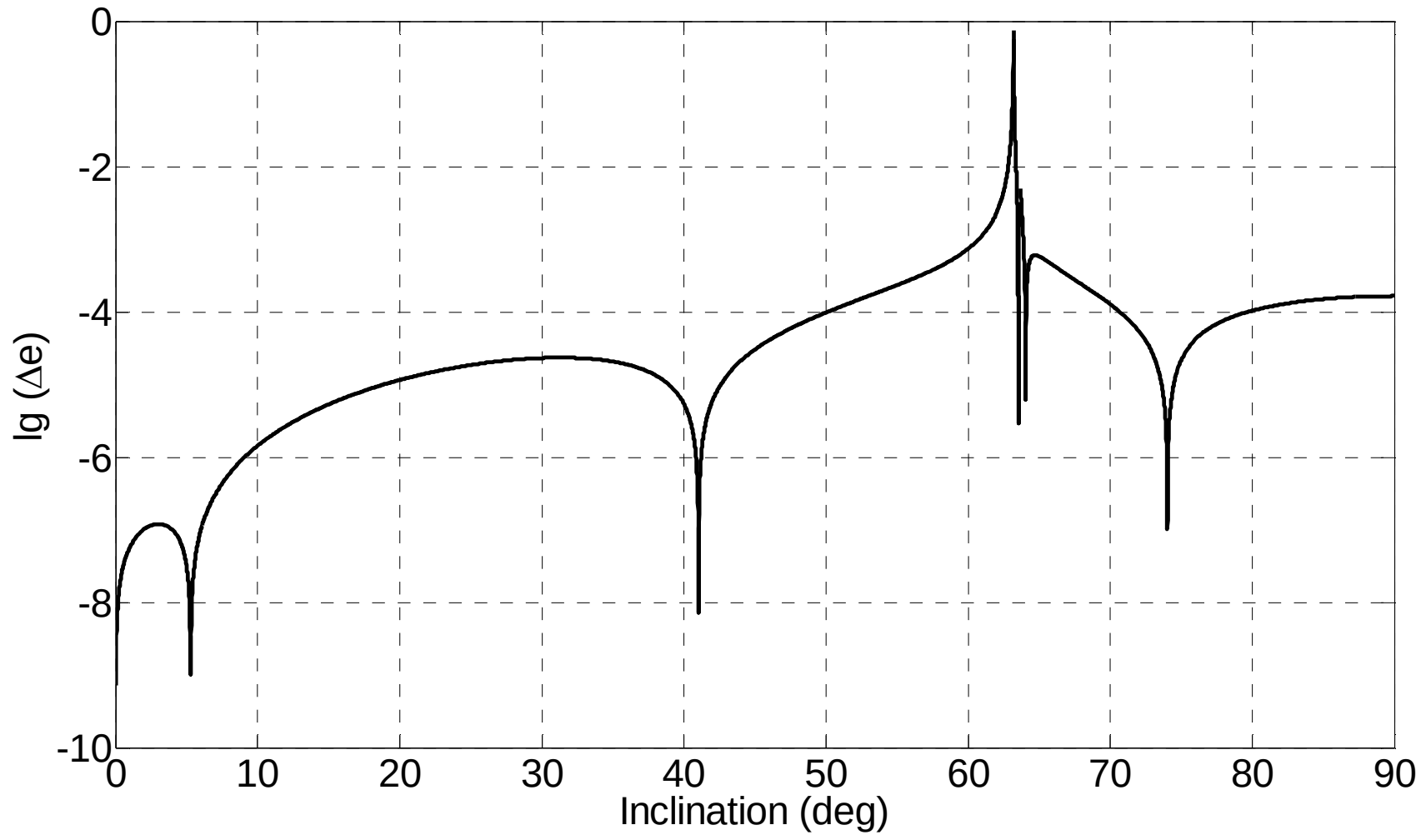

**Fig. 10** The logarithm of the differences in frozen eccentricities of the two different methods (Lagrangian formulations and Lie transforms) vs. inclinations for $a = 3597\ km$ in the zonal gravity field up to $J_9$.

One of Martian frozen orbits derived based on Lie transforms is tested in the $80 \times 80$ gravity field over one year. The set of orbital elements at the initial time are $a_0' = 3597\ km$ , $i_0' = 50°$ , $e_0' = 0.00736597$ , and $\omega_0' = 270°$ . The evolutions of the eccentricity and argument of pericentre can be presented in Fig. 11, and it shows a good property as an approximately frozen orbit. The eccentricity oscillates around $e_0'$, and the amplitude is about 0.0025; the argument of pericentre oscillates around $\omega_0'$, and the amplitude is about 20 degrees.

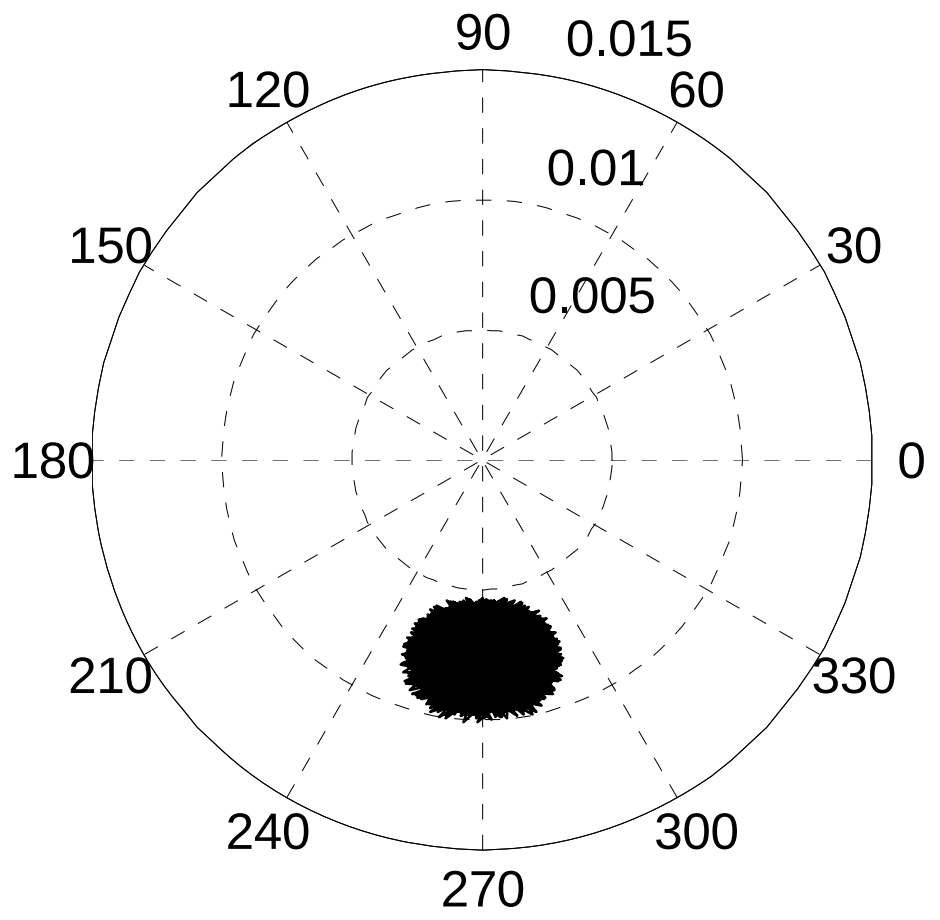

**Fig. 11** $e-\omega$ evolution over one year in the $80 \times 80$ gravity field for $a_0' = 3597\ km$ and $i_0' = 50°$. The initial condition is calculated based on Lagrangian formulations.

## 5 The Stability of Martian Frozen Orbits

By eliminating the cyclic coordinates, the problem up to $J_9$ can be reduced to two degrees of freedom, with a Routhian function

$$R = \frac{1}{2}\left(\dot{\rho}^2 + \dot{z}^2\right) - \frac{H^2}{2\rho^2} + \frac{\mu}{r}\left[1 - \sum_{l=2}^{9} J_l \left(\frac{R_e}{r}\right)^l P_l\left(\sin\varphi\right)\right], \tag{28}$$

where

$$\rho = r\cos\varphi;\ \ z = r\sin\varphi\,. \tag{29}$$

The variables $\rho$ and $z$ satisfy the Routh equation

$$\begin{aligned} \frac{\mathrm{d}}{\mathrm{d}t}\left(\frac{\partial R}{\partial \dot{\rho}}\right) - \frac{\partial R}{\partial \rho} = 0\,, \\ \frac{\mathrm{d}}{\mathrm{d}t}\left(\frac{\partial R}{\partial \dot{z}}\right) - \frac{\partial R}{\partial z} = 0\,. \end{aligned} \tag{30}$$

Using the notation

$$\begin{aligned} \mathbf{x} &= \left(\rho, \dot{\rho}, z, \dot{z}\right)^T, \\ \mathbf{f}\left(\mathbf{x}\right) &= \left(\dot{\rho}, \frac{\partial R}{\partial \rho}, \dot{z}, \frac{\partial R}{\partial z}\right)^T, \end{aligned} \tag{31}$$

the system (31) becomes

$$\dot{\mathbf{x}} = \mathbf{f}\left(\mathbf{x}\right). \tag{32}$$

Therefore, the fundamental solution matrix $\mathbf{\Phi}\left(t\right)$ yield the following

$$\dot{\mathbf{\Phi}}\left(t\right) = \mathbf{A}\left(t\right)\mathbf{\Phi}\left(t\right), \tag{33}$$

where

$$\mathbf{A}(t)=\left[\frac{\partial \mathbf{f}(\mathbf{x})}{\partial \mathbf{x}}\right]=\begin{bmatrix} 0 & 1 & 0 & 0 \\ \frac{\partial^2 R}{\partial \rho^2} & 0 & \frac{\partial^2 R}{\partial \rho \partial z} & 0 \\ 0 & 0 & 0 & 1 \\ \frac{\partial^2 R}{\partial z \partial \rho} & 0 & \frac{\partial^2 R}{\partial z^2} & 0 \end{bmatrix}. \tag{34}$$

The initial condition $\mathbf{\Phi}(0)$ used for the integration of Eq. (33) is the identity matrix. Then the linear stability of the system is determined by the eigenvalues of the monodromy matrix $\mathbf{\Phi}(T)$, where $T$ is the motion period.

For period orbits, the eigenvalues of $\mathbf{\Phi}(T)$ come in pairs. One pair are both equal to 1; another pair are real reciprocal of one another or two conjugate complex eigenvalues on the unit circle. Here, the stability index *k* is introduced to estimate the stability of the orbit

$$k = tr\left[\mathbf{\Phi}(T)\right]-2. \tag{35}$$

If $k > 2$ or $k < -2$, the orbit is unstable; if $-2 < k < 2$, the orbit is stable; and the conditions $k = \pm 2$ correspond to bifurcations.

The stability of Martian frozen orbits can be determined by the above-mentioned method. The general behavior of the stability index *k* of Martian frozen orbits for $a = 3597\ km$ calculated according to Lie transforms is shown in Fig. 12. It can be seen that for Martian frozen orbits over the range of inclinations $i \in [0, 90°]$, the stability indices are all close to 2. These orbits are approximately linearly stable. The simulation of one case far away from the critical inclination has been given in Fig. 11. It can be seen the oscillating amplitude of the eccentricity is about 0.0025, and the oscillating amplitude of the argument of pericentre is about 20 degrees. This case can be considered stable in engineering. Figure 13(a) presents the evolution of one frozen orbits close to the critical inclination with the initial orbital elements $a_0 = 3597\ km$, $i_0 = 63.34427571°$, $e_0 = 0.04$, and $\omega_0 = 90°$, and it shows that the oscillating amplitude of the eccentricity is about 0.01, and the oscillating amplitude of the argument of pericentre is about 40 degrees. In order to compare with the critical inclination cases given by (Liu et al. 2010), the same initial semimajor axis and eccentricity as Liu et al's are selected: $a_0 = 3897\ km$ and $e_0 = 0.1$. The corresponding inclination is derived based on Lie transforms: $63.26337591°$. The evolutions of the eccentricity and argument of pericentre is presented in Fig. 13(a). The oscillating amplitude of the eccentricity is about 0.004, and the oscillating amplitude of the argument of pericentre is about 7 degrees. It can be seen that the frozen property of Fig. 13(b) is better than those of Liu et al's Fig. 4 with $J_2$ and Fig. 6a with $J_2$ + $J_4$, and approaches that of Liu et al's Fig. 6b with $80 \times 80$ gravity calculated using the numerical method.

However, the stability analysis is limited to the zonal gravity model up to $J_9$. The stability analysis in the $80 \times 80$ gravity model will be investigated in the future.

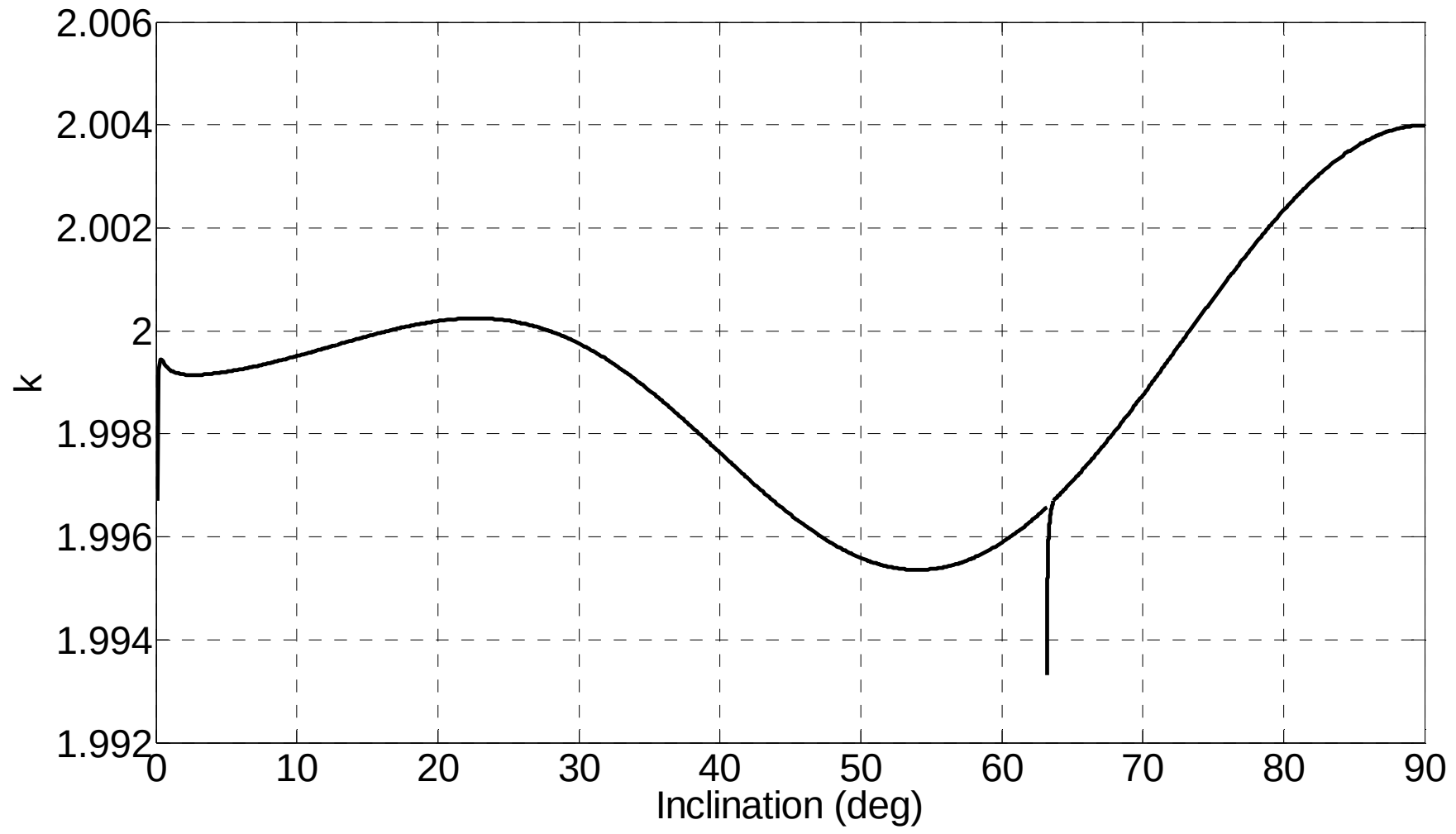

**Fig. 12** The stability index of Martian frozen orbits for $a = 3597\ km$ over the range of inclinations $i \in [0, 90°]$ in the zonal gravity field up to $J_9$.

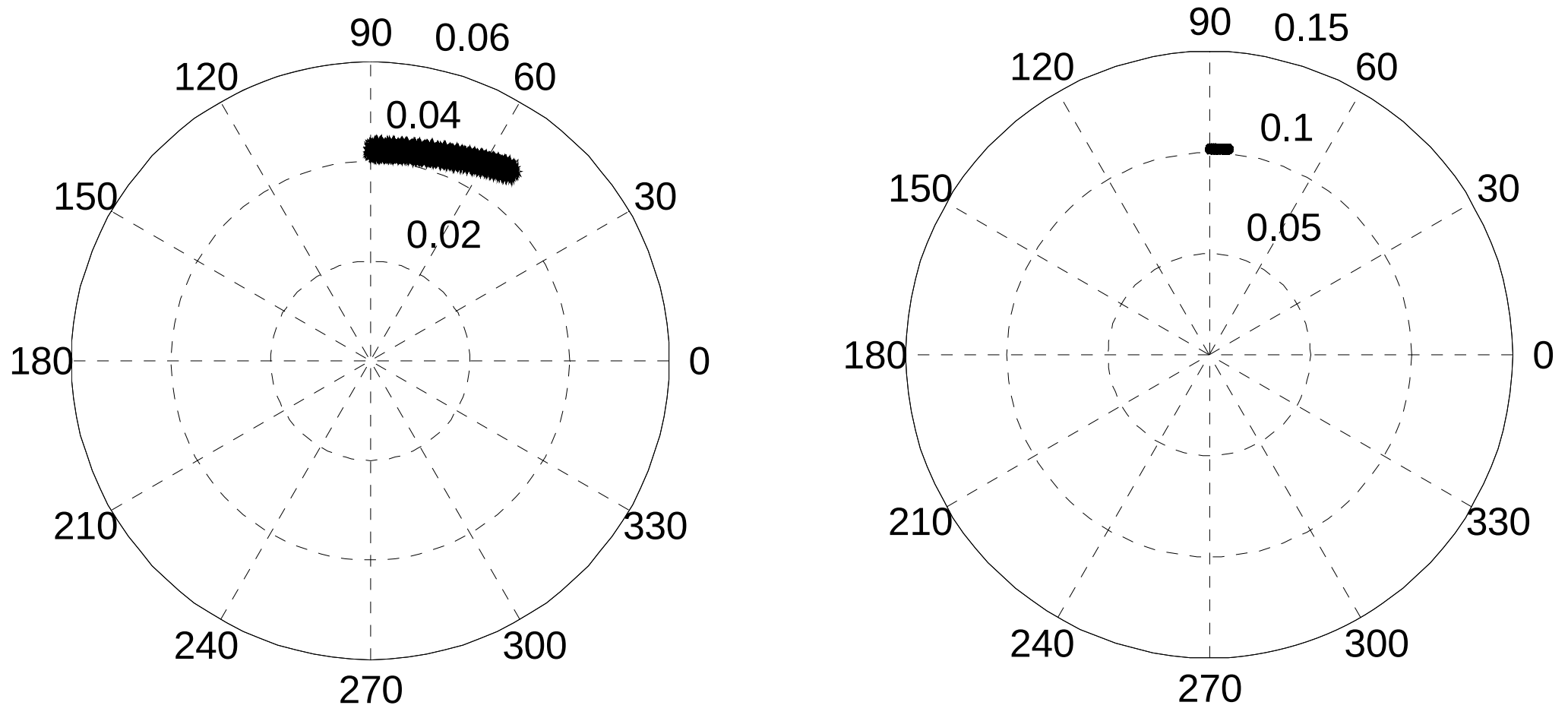

(**a**) (**b**)

**Fig. 13** $e - \omega$ evolutions over one year in the $80 \times 80$ gravity field. (**a**) Initial condition: $a_0 = 3597\ km$ , $e_0 = 0.04$ and $i_0 = 63.34427571°$ . (**b**) Initial condition: $a_0 = 3897\ km$ , $e_0 = 0.1$ and $i_0 = 63.26337591°$ .

## 6 Conclusions

This study provides two different analytical methods to analyze quasi-circular frozen orbits around Mars, and examine their basic nature. Based on Lagrangian formulations and Lie transforms, the constraint conditions for Martian frozen orbits are derived and the locations of Martian frozen orbits can be determined accurately. By comparison, the two methods are found well consistent with each other, and both discover four families of Martian frozen orbits: one family far away from the critical inclination with small eccentricity and the argument of pericentre equal to 90 degrees; two families far away from the critical inclination with small eccentricity and the argument of pericentre equal to 270 degrees; one family close to the critical inclination with the argument of pericentre equal to 90 degrees. The point with extremely small eccentricity is found at the boundary of two adjacent families of frozen orbits. With the variation of the polar component of the angular momentum, the evolution of the behavior of the equilibria in the phase plane, corresponding to locations of frozen orbits, is clarified. Simulations show that the two analytical methods can provide accurate enough results. The stability of Martian frozen orbits in the zonal gravity field up to $J_9$ is estimated based on the trace of the monodromy matrix. It is found that Martian frozen orbits over the range of inclinations $i \in [0, 90°]$ are approximately linearly stable. This paper provides initial conditions for numerical correction methods in the more complex models. The investigations of Martian frozen orbits in the realistic ephemeris model are in progress.

## Acknowledgments

This work was supported by the National Natural Science Foundation of China (No. 10832004 and No. 11072122).